\documentclass[12pt, a4paper]{article}
\pdfoutput=1
\usepackage{physics}
\usepackage{jheppub}
\usepackage{amsmath}
\usepackage{amssymb}
\usepackage{amsthm}
\usepackage{lmodern}
\usepackage{xcolor}
\usepackage{url}
\usepackage{cleveref}
\usepackage{hyperref}
\usepackage{graphicx}
\usepackage{subcaption}
\usepackage{mathrsfs}
\usepackage[utf8]{inputenc}
\usepackage[T1]{fontenc}
\hypersetup{
  colorlinks=true,
  citecolor=magenta,
  linkcolor=blue,
  urlcolor=violet,
  pdftitle={Upamanyu Moitra: Self-Similar Gravitational Dynamics,  Singularities and Criticality in 2D},
  pdfauthor={Upamanyu Moitra},
 }

\def\be{\begin{equation}}\def\ee{\end{equation}}
\newcommand{\baa}{\begin{equation}\begin{aligned}}
\newcommand{\ea}{\end{aligned}\end{equation}}

\title{Self-Similar Gravitational Dynamics,  Singularities and Criticality in 2D}

\author{Upamanyu Moitra}

\affiliation{International Centre for Theoretical Physics (ICTP),\\
Strada Costiera 11,  Trieste 34151, Italy}

\emailAdd{umoitra@ictp.it}

\abstract{We initiate a systematic study of continuously self-similar (CSS) gravitational dynamics in two dimensions,  motivated by critical phenomena observed in higher dimensional gravitational theories.  We consider CSS spacetimes admitting a homothetic Killing vector (HKV) field.  For a general two-dimensional gravitational theory coupled to a dilaton field and Maxwell field,  we find that the assumption of continuous self-similarity  determines the form of the dilaton coupling to the curvature.   Certain limits  produce two important classes of models,  one of which is closely related to two-dimensional target space string theory and the other being Liouville gravity.  The gauge field is shown to produce a shift in the dilaton potential strength. We consider static black hole solutions and find spacetimes with uncommon asymptotic behaviour.  We show the vacuum self-similar spacetimes to be special limits of the static solutions.  We add matter fields consistent with self-similarity (including a certain model of semi-classical gravity)  and write down the autonomous ordinary differential equations governing the gravitational dynamics.  Based on the phenomenon of finite-time blow-up in ODEs,  we argue that spacetime singularities are generic in our models.  We present qualitatively diverse results from analytical and numerical investigations regarding matter field collapse and singularities.  We find interesting hints of a Choptuik-like scaling law.}

\begin{document}
\maketitle

\section{Introduction}\label{sec-intro}

The general theory of relativity keeps springing surprises well past a century after its discovery.  Studies of different aspects of gravity continue to be rewarding.  The formation of black holes through gravitational collapse is one of the most striking predictions of general relativity.  For a better understanding of actual physical systems observed in Nature and of subtle conceptual issues,   the study of the \emph{dynamics} of gravitational systems is  very important.  As is true of most things important,  such studies are also very difficult. Even with assumptions that simplify a problem,  studying gravitational dynamics is a formidable task.  This fact has not prevented progress.  One can use several techniques, including numerical methods,  to glean important information about gravitational dynamics.

 One of the most surprising features in gravitational collapse was unravelled by numerical investigations by Choptuik \cite{Choptuik:1992jv}.  Choptuik considered a family of initial data for the spherically symmetric collapse (in $3+1$ dimensional flat spacetime ) of a massless scalar field,  characterised by a parameter $p$ (which, roughly speaking,  can be regarded as the ``strength'' of the initial data).  It turns out there exists a critical value $p^*$ (which is not universal) which governs whether or not the end state will contain a black hole.  If $p < p^*$, then the scalar field leaves behind an asymptotically flat spacetime with no black holes.  On the other hand,  if the strength $p$ exceeds $p^*$,  then a black hole is formed, whose mass near this threshold is governed by a scaling relation,
 \baa
M \sim (p - p^*)^\alpha, \label{chop}
 \ea
now commonly known as the \emph{Choptuik scaling law}.  While $p^*$ is non-universal,  the scaling exponent $\alpha$ is universal,  which was determined numerically.   It was further observed that around criticality,  the spacetime exhibits self-similarity.  It was also suggested that the critical solution itself corresponded to a naked singularity. {While the self-similarity observed in the original investigation was discrete in nature,  it was soon realised \cite{Evans:1994pj} that this similarity can be either discrete or continuous.} 
Behaviours of this kind (i.e., scaling,  universality) are ubiquitous in statistical mechanics.  Since the work of Choptuik,  there has been a considerable body of work in the exploration of gravitational critical phenomena.   An exposition of various aspects of critical behaviour in gravitational collapse can be found in the review \cite{Gundlach:2007gc}.

Since the solutions to the gravitational equations of motion show self-similarity near criticality,  a good place to start is to look for solutions with self-similarity as an \emph{input}.   The assumption of self-similarity is not overly restrictive, since it is observed in physical systems in a wide class of situations.  For a discussion on the importance of self-similarity in general relativity,  see the review \cite{Carr:1998at}, which also points out its relevance in astrophysics and cosmology.  From a practical point of view, the assumption of continuous self-similarity (CSS) accords an enormous simplification.   In this article,  we would concern ourselves with spacetimes admitting a homothetic Killing vector (HKV) field $\xi$, defined by,
\baa
\pounds_\xi g_{\mu \nu} = 2 g_{\mu \nu}\label{hkvdef},
\ea
where $\pounds_\xi$ denotes the Lie derivative.\footnote{The factor of $2$ on the right-hand-side of \cref{hkvdef} is conventional and corresponds to a normalisation of the vector field $\xi$.} The equations of motion become much simpler under the assumption of spacetimes admitting such a symmetry --- in our case (and in previous studies involving spherically symmetric spacetimes in higher dimensions),  the equations of motion turn into \emph{ordinary} differential equations (ODE) in a suitable self-similar variable,  which we call $x$. 

Some important insights about critical behaviour in gravitational collapse have come from studying ODEs of this kind.  Under a judicious set of transformations of the variables,  the ODEs in question become  autonomous differential equations (see \cite{Christodoulou:1994hg} and \cite{Brady:1994aq} for early works in this regard).    There exist methods and powerful theorems associated with solutions of autonomous ODEs under certain conditions --- this fact makes it easier to study the problem in question.     From the point of view of autonomous ODEs,  each integral curve in $x$ associated with a different initial condition corresponds to a different spacetime.  {The aforementioned} findings can be conveniently put in this language --- the final solutions  of scalar collapse with or without black holes represent two basins of attraction to which the integral curves flow depending on the initial condition $p$;  on the other hand,  the critical solution is a fixed point (saddle) of the ODEs which corresponds to a finely tuned initial condition $p=p^*$. Indeed,  as is common in a renormalisation group (RG) analysis,  one can read off various critical exponents from the eigenvalues obtained from a linear stability analysis around  the fixed point \cite{Koike:1995jm}.   The reader can refer to the review \cite{Gundlach:2007gc} and the references therein for more details; relatively recent applications of the assumption of homothety in different contexts include \cite{Zhang:2015rsa,  Rocha:2018lmv}.

The focus of the present article is on two-dimensional spacetimes with an HKV.  Before we describe our motivations for studying this problem in detail,  let us address an immediate concern some might have.  A homothetic Killing vector field defined by eq. \eqref{hkvdef} is a special case of a conformal Killing vector (CKV) field --- with the conformal pre-factor being a constant instead of an arbitrary function of spacetime coordinates.  As is well known,  \emph{every} two-dimensional metric  admits infinitely many CKVs.   Given this fact,  for any metric it is almost certain that there would exist an HKV.  However,  our interests are broader than the metric --- we consider two-dimensional dilaton gravity (coupled to matter) in which we have additional fields.  Since we want to find solutions exhibiting CSS,  we insist that the equations of motion transform appropriately under the action of the HKV --- we would thus need the transformation laws of the other fields under its action as well. Therefore,  the task of finding spacetimes with this symmetry is a non-trivial one and we will see that this symmetry is quite restrictive.

We now discuss the motivations for studying the problem in two dimensions.  First of all,  virtually all discussions of critical phenomena in gravity in four dimensions take place in the context of spherical symmetry.   The essential dynamics therefore takes place in a two-dimensional subspace. Under a spherically symmetric reduction of higher-dimensional gravity,  we end up precisely with dilaton gravity,     which offers a convenient setting to study such questions.  However,  two-dimensional dilaton gravity theories need not descend from higher dimensions --- they are quite interesting in themselves.  In fact,  we will consider dilaton gravity coupled to a Maxwell field.  When uplifted,  such a model can describe a wide variety of scenarios: rotating black holes, charged black holes in arbitrary dimensions with asymptotically flat or (A)dS spacetimes \cite{Moitra:2019bub},  black holes with toroidal geometry.   This work is motivated partly by the possibility of understanding critical behaviour in such scenarios.

Another motivation for this work is the prospect of an improved understanding of gravitational collapse and dynamical singularity  from a controlled microscopic perspective.  In two dimensions,  certain questions related to quantum gravity and string theory can be dealt with in a relatively tractable manner  --- matrix models are an important tool in this regard.   The resurgence of interest in two-dimensional gravity has led to several promising avenues in the recent years.  There is a considerable body of work on spacetime singularities in string theory, especially in the context of matrix models. See  for example,  the review \cite{Craps:2006yb} and the relatively recent work \cite{Das:2019cgl} and references therein.  (See also \cite{Moitra:2022glw}.) {The question of black hole (non-)formation is discussed in \cite{Karczmarek:2004bw}.}

Let us now describe our organisation of and the results obtained in this article. In  \textbf{Section \ref{sec-self}},  we describe our basic set-up of the ``pure'' or vacuum sector\footnote{In the sense that there are no bulk perturbative degrees of freedom.  This is in contrast with the situation when there are matter fields.}: metric, dilaton and a gauge field.  We consider a general dilaton gravity  action,  which can be reduced to a convenient form by a certain redefinition of the dilaton field.   With this redefinition,  there is a function $A(\phi)$ of the dilaton field $\phi$ which couples to the Ricci scalar.  We assume a very general transformation law for the dilaton field.  Demanding that some components of the equations of motion transform to themselves under the HKV allows us to fix the  form of the function $A(\phi)$.   For the most general form of this function $A(\phi)$,   it appears difficult to make further progress --- it turns out that under two different limits,  the form of $A(\phi)$ makes the analysis simple.  These two limits produce very well-known models.  The first model is closely related to  two-dimensional target space string theory \cite{Elitzur:1990ubs,  Mandal:1991tz,    Witten:1991yr},  we loosely refer to this class of models as the ``stringy'' models in the remainder of the paper.  The other model is nothing but Liouville gravity.  It is easy to fix the form of the remaining dilaton potential in both these models.   What about the gauge field? Owing to the fact that we are in two dimensions,  the homothety properties of the gauge field strength is completely determined by the remaining fields.  We find the curious result that the only effect of the gauge field is to  renormalise the strength of the dilaton potential.   This is a considerable simplification,  which allows us to eschew the gauge field altogether and work with an effective dilaton potential even when there is uncharged matter present.

In \textbf{Section \ref{sec-selsi}},  we first find the coordinate system which makes the study of CSS solutions very convenient --- with the particular choice we make,  all the relevant equations of motion become manifestly autonomous,  obviating the need for any further transformation of variables.  We first  solve the vacuum equations of motion under the CSS ansatz and find, rather unsurprisingly,  that the vacuum CSS solutions are some particular static solutions in disguise and also provide the explicit coordinate transformation between the static and CSS solutions.  We now add scalar matter to the theory.  We find some simple (yet conceptually relevant) matter models consistent with the CSS ansatz.  In particular, for the Liouville model,  we find that a certain semi-classical model is consistent with CSS --- which makes it an interesting object of study.  We write down the autonomous differential equations governing the dynamics for all the systems.  Here,  we observe a significant departure from related studies --- the previously mentioned linear stability analysis is no longer useful in our case, forcing us to look at the problem from a different angle.

In \textbf{Section \ref{sec-interlude}}, we use this fact (lack of utility of linearisation)  to our advantage to argue that in our models,  spacetime singularities are always formed on account of the phenomenon of \emph{finite-time blow-up},  a well-known feature of certain dynamical systems.  We illustrate our arguments with some simple dynamical systems.

In \textbf{Section \ref{sec-results}}, we present our results regarding the singularity and its nature in self-similar collapse with matter.  In certain simple cases,  we are able to make statements analytically.  In most cases, however, we have to take recourse to high-precision numerics to make proper statements about the solutions.  (We emphasise that, within the models,  the solutions discussed are fully back-reacted and no approximation like the probe limit has been used anywhere.)  We find very interesting results about the singularity in all these models.  In particular,  we find numerical evidence of a behaviour analogous to Choptuik scaling.    Virtually all the singularities  in the dynamical situation turn out to be null or spacelike,  but not timelike.  Even when a timelike singularity is present initially,  it is rendered unstable by the dynamics. The results are thus pleasing vis-à-vis the weak cosmic censorship conjecture.  (See \cite{Moitra:2020ojo} and \cite{Moitra:2021eom} for discussions on the \emph{strong} cosmic censorship conjecture in two dimensions.)

We conclude this article with a discussion on many exciting future directions in \textbf{Section \ref{sec-concl}}.   

In \textbf{Appendix \ref{sec-static}, } we write down the static solutions corresponding to the actions obtained in Section \ref{sec-self}. We make a detailed study of the black hole solutions for different values of the parameters in the theory.  Although these solutions are not self-similar in general,  the numerical analysis in Section \ref{sec-results} is informed by the results of this appendix.  We find, rather surprisingly,  a class of spacetimes which have a similar asymptotic causal structure as AdS spacetimes even though the curvature is vanishingly small asymptotically.  Such spacetimes of higher dimensionality have been observed previously \cite{Chan:1995fr}. To the best of my knowledge,  this is the first time explicit examples of such unusual asymptotics have been constructed in two dimensions.  {Since this appendix is quite useful in building the intuition for the analysis in the main body of this article,  some readers might want to study this appendix before starting Section \ref{sec-selsi}.}

An existing definition of quasi-local mass,  which is useful in the main text,  is discussed in \textbf{Appendix \ref{app-qlm}}.

Since some of our two-dimensional solutions asymptote to spacetimes which look similar to AdS spacetimes,  this also raises the exciting possibility that we may gain insight about self-similar spacetimes using some holographic model.  A constant non-zero value of the scalar curvature $R$ is inconsistent with the existence of an HKV \eqref{hkvdef}.  The authors of \cite{Pretorius:2000yu} found continuously self-similar behaviour near criticality,  which was \emph{not} associated with   an HKV.   Diverse investigations in the context of higher dimensional AdS spacetimes include \cite{Birmingham:1999yt,  Birmingham:2001hc, Bhattacharyya:2009uu,  Chesler:2019ozd,  Emparan:2021ewh}. 
As an example of the interesting possibilities of self-similar gravitational collapse involving holography,   such a collapse was related to high-energy scattering in QCD in \cite{Alvarez-Gaume:2008emf}. 
It would be interesting indeed if the bulk dynamics explored in this article could be phrased entirely in terms of some boundary degrees of freedom.

Before we proceed further,  let us mention some other relevant references.  See \cite{Wald:1997wa} for an early review of weak cosmic censorship in the context of gravitational collapse (see also \cite{Joshi:2008zz}).   Questions related to Choptuik scaling and critical collapse in two spacetime dimensions have been considered in a few articles \cite{Strominger:1993tt,  Peleg:1996ce, Frolov:2006is, Dhar:2018pii}.   Such investigations differ significantly from the one presented here --- the assumption of continuous self-similarity plays the central role in our discussions.  In the previous works, self-similarity, if present, was discrete in nature.  Indeed,  in a specific model, there were arguments \emph{against} continuous self-similarity \cite{Chiba:1997ex}.  Instead of starting with some specific models, we begin systematically by asking the question: what models are allowed at all? 
Furthermore,  many such specific models are subsumed as special cases of the broad class of models described in this article.  We hope the work presented in this article will be useful in further investigations of critical collapse. 

\section{Self-Similarity in Two Dimensions}\label{sec-self}

We start our discussion with the  ``pure'' or vacuum sector, described by a general two-derivative action
with the metric,  dilaton and Maxwell field\footnote{The gauge potential is of course $A_\mu$ and the field strength is $F_{\mu \nu} = \partial_\mu A_\nu -  \partial_\nu A_\mu$.  Though we have used the same letter for the gauge potential $A_\mu$ and the function $A(\phi)$,  the difference should be obvious from the context.  Actually,  we have  not used the gauge potential explicitly anywhere in the article. },
\baa
S = \frac{1}{16 \pi G_2} \int \dd[2]{x} \, \sqrt{-g} A(\phi) \bqty{ R + \gamma                    (\nabla \phi)^2 + V(\phi) - G(\phi) F_{\mu \nu} F^{\mu \nu}   }, \label{genac}
\ea
and examine what restrictions are imposed by the criterion of self-similarity.  The reader might have noticed that we have chosen the same function of the dilaton field $A(\phi)$ before the Ricci scalar and the dilaton kinetic term.  In the most general possible action,  the coefficient of the Ricci scalar term would be different from the one of the kinetic term.  However,  we can redefine the field $\phi$ to make the two terms equal.

We could also eliminate the kinetic term for the dilaton by making a dilaton-dependent Weyl transformation of the metric.  
Since we wish to study the singularity properties and also connect the two-dimensional physics to higher dimensions,  we work in a fixed Weyl frame throughout this article,  which makes the connection more transparent.
We shall work with some fixed $A(\phi)$  --- we keep the coefficient $\gamma$ as an independent parameter as different values of $\gamma$ illustrate qualitatively different aspects of the physics, as will be apparent below.    See also \cite{Lechtenfeld:1992rt} for related comments.

The equation of motion obtained by varying the scalar $\phi$ is given by,
\baa
R - \gamma ( \nabla \phi )^2 - \frac{2\gamma A(\phi)}{A'(\phi)} \nabla^2 \phi+\pqty{V(\phi) +  \frac{V'(\phi) A(\phi)}{A'(\phi)} } - \pqty{ G(\phi) + \frac{G'(\phi) A(\phi) }{A'(\phi)}  } F^2 =0. \label{phiom}
\ea
The  equation of motion obtained by varying the (inverse) metric is given by,
\baa
&g_{\mu \nu} A'(\phi) \nabla^2 \phi + g_{\mu \nu} A''(\phi) (\nabla \phi)^2 - A''(\phi) \nabla_\mu \phi \nabla_\nu \phi - A'(\phi) \nabla_\mu \nabla_\nu \phi \\
&\quad + \gamma       A(\phi) \pqty{ \nabla_\mu \phi \nabla_\nu \phi - \frac12  (\nabla \phi)^2 g_{\mu \nu} } - \frac12 g_{\mu \nu} A(\phi ) V(\phi) \\
&\quad - 2A(\phi) G(\phi)\pqty{  F_{\mu \rho}  F_{\nu} {}^\rho - \frac14 g_{\mu \nu} F^2} = 0. \label{meom1}
\ea
Its trace gives,
\baa
A'(\phi)  \nabla^2 \phi + A''(\phi) (\nabla \phi)^2  - A(\phi) V(\phi) - A(\phi) G(\phi) F^2 = 0. \label{treom}
\ea

\subsection{Most General Curvature Coupling}

Since we are looking for self-similar solutions,  we demand that under the action of the HKV,  the equations of motion \eqref{phiom} and \eqref{treom} should transform to themselves, up to some overall multiplicative factor.

Under the homothety transformation \eqref{hkvdef}, the Ricci scalar transforms with weight $(-2)$,
\begin{equation}
\pounds_\xi R = - 2 R.  \label{rictrans}
\end{equation}

Under our assumptions,  we will see that we will be able to constrain the form of $A(\phi)$ by simply considering the transformation properties of $(\nabla \phi)^2$ and $\nabla^2 \phi$.

Let us  note that for any HKV $\xi$, 
\baa
\pounds_\xi \nabla_\mu \phi =  \nabla_\mu \pounds_\xi \phi, \label{hvgrad}
\ea
and,
\baa
\pounds_\xi \nabla^2 \phi =  \nabla^2 \pounds_\xi \phi - 2 \nabla^2 \phi. \label{hvlap}
\ea

Let us make the reasonable assumption under the Lie action of the HKV,  the scalar field transforms to a function of itself (which is not a function of the metric or the gauge field)\footnote{One could possibly consider here derivatives of $\phi$ --- but we would necessarily have to include the metric to soak up the indices.  The ensuing analysis would be rather complicated.}
\baa
\pounds_\xi \phi = f(\phi). \label{asphi}
\ea

To complete the discussion we would also need to consider the homothety transformation properties of the gauge field.  However,   since our immediate goal is to determine the form of the function $A(\phi)$, we can ignore the gauge field term for the moment.  It will turn out that the gauge field homothetic properties will be uniquely determined by those for the metric and the dilaton.

Under the assumed conditions,  only the first term of the LHS of eq. \eqref{phiom} generates the Ricci scalar under the HKV action.  By the property \eqref{rictrans},  we can thus say that for self-similar solutions,  the homothety action transforms the LHS of \eqref{phiom} to $(-2)$ times itself.

We see for the terms involving derivatives of $\phi$,   using eqs.  \eqref{hvgrad},  \eqref{hvlap} and \eqref{asphi},
\baa
\pounds_\xi \pqty{- \gamma (\nabla \phi)^2 } &= (-2)(-\gamma (\nabla \phi)^2)  - 2 \gamma f'(\phi) (\nabla \phi)^2, \\
\pounds_\xi \bqty{- \frac{2\gamma A(\phi)}{A'(\phi)} \nabla^2 \phi } &= (-2) \bqty{- \frac{2\gamma A(\phi)}{A'(\phi)} \nabla^2 \phi }  - 2 \gamma \frac{A(\phi)}{A'(\phi)} f''(\phi) (\nabla \phi)^2 \\
&\quad - 2 \gamma \dv{\phi} (\frac{A(\phi)f(\phi)}{A'(\phi) } ) \nabla^2 \phi. \label{lst1}
\ea

Since no other terms can generate derivatives of $\phi$ under the homothety transformation rules,   for \cref{phiom} to be covariant under a homothety transformation,  we obtain  the following condition on the derivatives of $\phi$,
\baa
-2\gamma \pqty{ f'(\phi) + \frac{A(\phi)		}{A'(\phi)} f''(\phi) } (\nabla \phi)^2  - 2 \gamma \dv{\phi} (\frac{A(\phi)f(\phi)}{A'(\phi) } ) \nabla^2 \phi  = 0. \label{lst2}
\ea
The coefficients of these terms should vanish independently and we   immediately obtain,
\baa
f(\phi) = a_1 \frac{A'(\phi)}{A(\phi)}, \label{aten}
\ea
and,
\baa
f'(\phi) = \frac{a_2}{A(\phi)},\label{ael}
\ea
where $a_{1,2}$ are constants.

Let us see if homothety action on the derivative terms of \eqref{treom} produces more constraints.  Since the coefficients of $(\nabla \phi)^2$ and $\nabla^2 \phi$ are  $A''(\phi)$ and $A'(\phi)$ respectively,  we insist that under a homothety transformation,  ratio of the coefficient of $(\nabla \phi)^2$ to that of $(\nabla^2 \phi)$ must be the same as $A''(\phi) / A'(\phi)$. 

We see,
\baa
\pounds_\xi \bqty{ A''(\phi) (\nabla \phi)^2 } =  \bqty{ -2 A''(\phi) + 2A''(\phi) f' (\phi)  + A'''(\phi) f(\phi) }(\nabla \phi)^2,  \label{etof}
\ea 
and,
\baa
\pounds_\xi \bqty{ A'(\phi) \nabla^2 \phi } = \bqty{-2 A'(\phi)  + A''(\phi) f(\phi) + A'(\phi) f'(\phi) } \nabla^2 \phi + A'(\phi) f''(\phi) (\nabla \phi)^2   .  \label{etofn}
\ea 

Therefore,  on acting with $\pounds_\xi$ on equation \eqref{treom},  the resulting coefficient of $(\nabla \phi)^2$ is given by,
\baa
A'''f + 2 A'' f' + A' f'' - 2 A'' = (A'f - 2 A)'', \label{etosxn}
\ea
and that of $\nabla^2 \phi$ is given by,
\baa
A'' f + A' f' - 2 A '  = (A'f - 2A)'. \label{etosvn}
\ea
Therefore,  we must have,
\baa
\frac{(A'f - 2A)''}{(A'f - 2A)'} = \frac{A''}{A'}. \label{etoegn}
\ea

This determines the form of $f(\phi)$ to be,
\baa
 f (\phi ) =\frac{ a_3 A (\phi) + a_4}{A' (\phi)} \label{frel2},
\ea
with new constants $a_{3,4}$.
Substituting eq. \eqref{aten} in eq. \eqref{frel2} and solving the resulting differential equation,  we obtain the general solution for $A(\phi)$,
\baa
A(\phi) =  \frac{a_4}{a_3} \sinh^2 \pqty{ \frac12 \sqrt{\frac{a_3}{a_1}} (\phi - \phi_0)  }, \label{aphifin}
\ea
where $\phi_0$ is an arbitrary constant of integration.  Recall from eq. \eqref{genac} that $A(\phi)$ couples to the Ricci scalar and hence must be positive.  We could also insist on this coupling being monotonic in $\phi$. These conditions would imply that $a_1$, $a_3$ and $a_4$ are of the same sign.

As a check, we find that the form of $f(\phi)$ determined from \eqref{aten}  is consistent with the behaviour from \eqref{ael},  with the constant $a_2$ being determined to be $-a_4/2$.

Although we have obtained the form of $A(\phi)$ consistent with the assumed homothety \eqref{hkvdef} and the assumption \eqref{asphi}, it seems difficult to proceed further with the general form \eqref{aphifin}.  We will instead focus on a subset of the most general solution in the theory space,  which are still quite generic,  and we find that these models are rather well-known. 

\subsection{Models Related to 2D Target Space String Theory}

One of the simplest results emerges when we set $a_4 = 0$ in eq. \eqref{frel2}.  The $a_4 \to 0$ limit of eq. \eqref{aphifin} is trivial so we examine this case separately.  When $a_4 = 0$,  eqs. \eqref{frel2} and \eqref{aten} together imply that $A'(\phi)/A(\phi)$ must be a constant. Since $A(\phi)$ is real,  we  reach the conclusion that $A(\phi)$ is an exponential function in $\phi$.  We can use our freedom to rescale the Newton constant and the coefficient $\gamma$ to set
\baa
A(\phi) = \exp(-2\phi). \label{aphi}
\ea
This form (involving the Ricci scalar and the dilaton kinetic term) is exactly the same as two-dimensional target space string theory.  For string theory,  the value of the coupling $\gamma$ is 4.  

Once we have fixed the scale and origin of $\phi$,  we can now fix $f(\phi)$ to be some constant.  Following the conventions in the literature,  we take the Lie derivative of the scalar field to be,
\baa
\pounds_\xi \phi = - \kappa.
\label{ldevkap}
\ea

We would now like to determine the forms of the potential $V(\phi)$ and the ``gauge coupling'' $G(\phi)$.  The Maxwell equations are given by,
\baa
\nabla_\mu ( A(\phi) G(\phi) F^{\mu \nu} ) = 0. \label{feom}
\ea
Since the discussion concerns a two-dimensional spacetime,  we must have,
\baa
A(\phi) G(\phi) F_{\mu \nu} = Q \sqrt{-g} \epsilon_{\mu \nu},  \label{feom1}
\ea 
where $\frac12 \sqrt{-g}\, \epsilon_{\mu \nu} \, \dd x^\mu \wedge \dd x^\nu$ is the volume form and $Q$ is a constant,  which is  the charge carried by the system.  Using \cref{feom1} and the properties of the volume form,  we readily obtain,
\baa
F_{\mu \rho} F_{\nu} {}^\rho &= - \frac{Q^2}{A^2  G^2  } g_{\mu \nu}, \\
F^2  &= - \frac{2Q^2}{A^2 G^2 }. \label{feom2}
\ea

Inserting \cref{feom2} in \cref{treom} and using \cref{aphi} we obtain,
\baa
- \nabla^2 \phi + 2(\nabla \phi)^2  -   \frac12 V(\phi) + \frac{Q^2 e^{4\phi}}{ G(\phi) } = 0.  \label{treom2}
\ea
If we demand that the homothety properties be independent of the some specific solutions to the equation of motion, we could analyse the properties of the two functions appearing in \cref{treom2} independently.   Since both $\nabla^2 \phi$ and $(\nabla \phi)^2$ transform under homothety with weight $-2$,  we must demand that,
\baa
\pounds_\xi V(\phi) = -2 V(\phi).  \label{lvmtv}
\ea 
This implies that,
\baa
- \kappa V'(\phi) =  -2 V(\phi), \label{lvmtv2}
\ea
which constrains the potential to be of the form,
\baa
V (\phi ) = V_0 \exp ( \frac{2  \phi}{\kappa}).  \label{vpdef}
\ea
By a similar argument, we must have $e^{4\phi} / G(\phi) = e^{2\phi/\kappa}$ and thus, we obtain the required functional form for $G(\phi)$,
\baa
G(\phi) = \exp (4 \phi- \frac{2  \phi}{\kappa}).  \label{Gdef}
\ea

Before proceeding further,  let us determine the homothety property of the field strength tensor $F_{\mu \nu}$ and check for consistency. We have from \cref{feom1},
\baa
\exp(2 \phi - \frac{2\phi}{\kappa} )F_{\mu \nu} = Q \sqrt{-g} \epsilon_{\mu \nu}. \label{feom3}
\ea
We note that under the homothety transformation $\sqrt{-g} \epsilon_{\mu \nu}$ transforms to itself with weight 2,
\baa
\pounds_\xi \sqrt{-g} \epsilon_{\mu \nu} = 2 \sqrt{-g}\epsilon_{\mu \nu}. \label{htvolf}
\ea
Therefore,  the left hand side must transform with weight 2 as well.  Since the quantity $\exp(2 \phi (1-1/\kappa) )$ transforms with weight $2 (1-\kappa)$ (by \eqref{ldevkap}),  the weight of $F_{\mu \nu}$ is uniquely determined to be $2 \kappa$:
\baa
\pounds_\xi F_{\mu \nu} =2 \kappa F_{\mu \nu}. \label{htfmn}
\ea
Therefore,  from \eqref{treom},  we see that $G(\phi) F^2$ must transform with weight $(-2)$.  Noting that $F^2$ has two factors of the inverse metric and two factors of the field strength $F_{\mu \nu}$,  we find that the weight of this quantity is,
\baa
\pqty{ 4 - \frac{2}{\kappa} } (-\kappa) +2 (-2) + 2 (2 \kappa) = -2. \label{kapide}
\ea
Thus the homothety properties are consistent, as they must be.

Let us write down the final equations of motion. The $\phi$ equation of motion is given by,
\baa
R - \gamma ( \nabla \phi )^2 + \gamma \nabla^2 \phi+ (V_0 - 2 Q^2)  \pqty{1-\frac{1}{\kappa }} \exp(  \frac{2  \phi}{\kappa} )  =0.  \label{phiom1}
\ea
On the other hand,
\baa
&2  \nabla_\mu \nabla_\nu \phi - 2 g_{\mu \nu}  \nabla^2 \phi + \frac{8-\gamma}{2} g_{\mu \nu}  (\nabla \phi)^2 + (\gamma -4) \nabla_\mu \phi \nabla_\nu \phi  - \frac12 g_{\mu \nu} e^{  \frac{2  \phi}{\kappa} }  ( V_0 -  2Q^2 )= 0. \label{meom2}
\ea
It is thus easy to see that the effect of the gauge field is to additively shift the value of potential strength and therefore,  as far the source-free equations are concerned, we could work with only an effective potential strength
\baa
V_{\mathrm{eff}}  \equiv V_0 - 2 Q^2, \label{vefdef}
\ea 
and no gauge field. 

\subsection{Liouville Gravity}\label{subsec-lioset}

We can examine another case,  which is given by $a_3 = 0$.  In this case, the solution for $A(\phi)$ is given by,
\baa
A(\phi) = \frac{a_4}{4a_1} (\phi - \phi_0)^2. \label{lio1}
\ea
Rescaling and shifting $\phi$,  we can take 
\baa 
A(\phi) = \phi^2. \label{liod}
\ea
This is actually nothing but Liouville gravity in disguise. The simple field redefinition,
\baa
\phi^2 = \sigma_0+ \sigma , \label{psfred}
\ea
yields the action for Liouville gravity,
\baa
S = \frac{1}{16\pi G_2} \int \dd[2]{x} \sqrt{-g} \pqty{ (\sigma_0+ \sigma)  R - ( \nabla \sigma)^2 + V(\sigma) - G(\sigma) F^2 }. \label{lioac}
\ea
Here, $\sigma_0$ is a constant which we have separated by hand from the dynamical field $\sigma$ --- this artificial splitting has been done for later convenience when we discuss the black hole solutions and collapse.

Note that in this case,  we can absorb $\gamma$ before the kinetic term by a redefinition of the $\sigma$ field and the Newton constant $G_2$.  Note that we have chosen a particular sign for the kinetic term for clearly establishing the connection between static black holes (see Appendix \ref{sec-static}) in this model  and those of the stringy model.  

The equations of motion are given by,
\baa
R + 2 \nabla^2 \sigma + V'(\sigma) - G'(\sigma) F^2 &= 0,  \label{lieom}
\ea
and
\baa
g_{\mu \nu} \nabla^2 \sigma - \nabla_\mu \nabla_\nu \sigma &-\pqty{  \nabla_\mu \sigma \nabla_\nu \sigma - \frac12 (\nabla \sigma)^2 g_{\mu \nu} } \\
&- \frac12 g_{\mu \nu} V  - 2 G \pqty{  F_{\mu \rho}  F_{\nu} {}^\rho - \frac14 g_{\mu \nu} F^2} =0. \label{liogeom}
\ea
We now have,
\baa
\pounds_\xi \sigma = - \kappa, \label{liohomota}
\ea
and similar considerations as in the previous subsection fix $V(\sigma)$ and $G(\sigma)$ to take the forms,
\baa
V(\sigma) &= V_0  \exp(  \frac{2\sigma}{\kappa} ),  \\
G(\sigma) &=  \exp(  -\frac{2\sigma}{\kappa} ). \label{vsigdef}
\ea

As before,   if we insert the solution to the Maxwell equations,  the effective potential strength is shifted by the charge and we obtain the same relation as eq.  \eqref{vefdef},  i.e.,
$$
V_{\mathrm{eff}} = V_0 - 2 Q^2.
$$
In the remainder of this article,  we work with an effective potential strength and no gauge field.

\section{Self-Similar Solutions:  Setting the Stage}\label{sec-selsi}

\subsection{The Coordinate System}\label{subsec-coord}

We need to work with a coordinate system well-adapted to the problem at hand --- namely,  describing the self-similar behaviour of gravitating systems.   In higher dimensions, typically the retarded or advanced Bondi gauge is particularly useful.  In studies in higher dimensions (see, for example, \cite{Brady:1994aq}) an ansatz like the following is used,
\baa
\dd s^2 = -g \qty( \frac{r}{v}) \bar{g} \qty( \frac{r}{v}) \dd{v}^2 + 2 \bar{g} \qty( \frac{r}{v})  \dd{v} \dd{r} + r^2 \, \dd \Omega_2^2. \label{brady}
\ea
Here $v$ is an advanced null coordinate,  and one has to consider a patch of spacetime,  say $v>0$.
This metric transforms appropriately under the action of the homothetic Killing vector field,
\baa
\xi = v \pdv{v} + r \pdv{r}. \label{xiru}
\ea
In other words, under the transformation $v \to \lambda v \, \, r \to \lambda r$, the line element transforms as $\dd s^2 \to \lambda^2 \, \dd s^2$.  Since we are interested in a two-dimensional spacetime,  we could start with the line element by simply omitting the transverse part of the metric \eqref{brady}.  We thus have the initial metric ansatz,
\baa
\dd s^2 =  \bar{g} \qty( \frac{r}{v})  \bqty{ -g \qty( \frac{r}{v}) \dd v^2 + 2\,   \dd v \, \dd r }.  \label{dsru1}
\ea
It turns out that we can  simplify this ansatz considerably by using appropriate coordinate transformations and get rid of the Weyl factor $\bar{g}$ altogether.  This will aid the later calculations considerably.  Let us first define,
\baa
r = y v.  \label{ryu1}
\ea
In the $(y, v)$-coordinate system the metric takes the form,
\baa
\dd s^2 = \bqty{ -\bar{g} (y) g(y) + 2 y \bar{g}(y) } \dd v^2 + 2 v \bar g(y)  \dd{y}\dd{v}.  \label{dsyu}
\ea
Let us define a new coordinate $x$,
\baa
x = \int^y \dd{y'} \bar g(y'),  \label{xinty}
\ea
which can be inverted to write $y$ as a function of $x$,
\baa
y = y(x). \label{yyx}
\ea
In terms of this new coordinate, the line element can be written as,
\baa
\dd s^2 =  \bqty{ -\bar{g} (y (x) ) g(y(x) ) + 2 y(x) \bar{g}(y(x) ) } \dd v^2 + 2 v  \dd{x}\dd{v}. \label{dsxu}
\ea
We retain the old $v$-coordinate,  and define a new ``radial'' coordinate $\rho$, 
\baa
x =  \frac{\rho}{v}. \label{xdef}
\ea
The line element then reads,
\baa
\dd s^2 =  -\bqty{ \bar{g} (y (x) ) g(y(x) ) - 2 y(x) \bar{g}(y(x) ) + 2 x } \dd v^2 + 2  \dd{\rho }\dd{v}.  \label{dsRU1}
\ea
Defining $F(x) \equiv  \bar{g} (y (x) ) g(y(x) ) - 2 y(x) \bar{g}(y(x) )$, we can write the simplified self-similar ansatz as,
\baa
\dd s^2 = -( F(x) + 2x) \dd{v}^2 + 2 \dd{v}\dd{\rho},  \label{cssan}
\ea
which we shall use in the rest of the article. We emphasise at this point that $(\rho,v)$  coordinates used here are different from the static coordinates either in the Schwarzschild gauge or in the Eddington-Finkelstein gauge used below.

Note that we have done away with one function here in deriving the final form starting from \eqref{dsru1}.  This can be done in the $(r,v)$ subspace in higher dimensions as well, where the ``extra'' undetermined function gets transferred to the transverse part of the metric.  However,  in two dimensions there is no transverse part of the metric.  When the dilaton gravity arises from dimensional reduction of higher-dimensional gravity, it is the dilaton that  plays role of the transverse sphere. Note that the ansatz above has the same homothety property as before,
\baa
\xi = v \pdv{v} + \rho \pdv{\rho}. \label{xiUR}
\ea
This coordinate chart has the advantage of being well-adapted to describe gravitational collapse.

We note that the Ricci scalar assumes a particularly simple form in this coordinate chart,
\baa
R = - \frac{F''( x )}{v^2}. \label{ricUR}
\ea
We will often use the $x$ variable in our investigations to describe our solutions.  In trying to find the self-similar solutions, we will adopt exactly the same strategy as for the static solutions discussed in Appendix \ref{sec-static}.  We use the ansatz \eqref{cssan}  and insert it in the equations of motion to solve for it.  We would also need an ansatz for the dilaton (and other matter fields).  

Recalling that in all cases of interest,  the action of the vector field \eqref{xiUR} on the scalar $\phi$ (or the Liouville field $\sigma$) is the constant $(-\kappa)$,  we immediately solve for $\phi$,
\baa
\phi (\rho , v) = \Phi \pqty{x} - \kappa \log v.  \label{cssphi},
\ea
where $\Phi(x)$ is a function to be determined from the equations of motion.

Let us mention briefly here that we could have also worked with retarded null coordinate $u$,  which could be obtained by the substitution $v \to -u$ in the formul\ae\ above.  In this case,  one describes the patch $u<0$.  The form of all the equations of motion remain unchanged under this transformation --- therefore the solutions we obtain later can also be adapted to this coordinate. 
This fact can be used for patching together different regions of spacetime.
The advanced null coordinate seems  to be suitable for the initial value problem we are interested in.

\subsection{Why Matter Matters: The Vacuum Self-Similar Solutions}\label{subsec-matmat}

The most obvious thing to do at first would be to consider the vacuum equations and solve them using the using the self-similar ans\"{a}tze  \eqref{cssan} and \eqref{cssphi}. The procedure is roughly the same as that of finding the static solutions; see Appendix \ref{sec-static}.   We do not go through the steps in detail at this point --- some explicit details are given later when we deal with the system in the presence of additional matter.  The results in this subsection can be treated as a limiting case for the more general scenarios discussed later.

Let us then write down the solutions for the metric function $F(x)$ and the dilaton function $\Phi (x)$. For the stringy system with $\gamma < 4$,  the solutions are given by,
\baa
F(x) =   \frac{(4-\gamma)^2 \kappa  V_{\mathrm{eff}}}{4 ((8 - \gamma) \kappa -4)} x^{2 - \frac{4}{(4-\gamma ) \kappa } } - (4-\gamma) \kappa x,  
\label{fxglf}
\ea
and,
\baa
\Phi (x) = - \frac{2}{4-\gamma}  \log x.  \label{phixglf}
\ea
Before we proceed further,  it is important to address one important question -- whether the solution we just found represents a genuinely new solution or it is a familiar solution in a different guise.  It should be no surprise that the latter  turns out to be correct.  To see this, let us write down the static solution \eqref{bfphi1}  with $r_h \to 0$  in the ingoing Eddington-Finkelstein coordinate chart (we use $\bar{v}$ as the null coordinate to distinguish it from the $v$ used in the self-similar ansatz). The line element reads,
\baa
\dd s^2 = -\frac{(4-\gamma)^2 \kappa  V_{\mathrm{eff}}}{4 ((8 - \gamma) \kappa -4)} r^{2 - \frac{4}{(4-\gamma ) \kappa } }  \,\dd \bar{v}^2 + 2 \, \dd \bar{v} \, \dd r.  \label{leefrst}
\ea
The coordinate transformation that brings this line element and dilaton into the aforementioned forms is,
\baa
r &= \rho v^{ \frac{(4-\gamma) \kappa}{2}-1 }, \\
\bar{v} &= \frac{1}{2-\frac{1}{2} (4-\gamma ) \kappa} v^{2-\frac{1}{2} (4-\gamma ) \kappa}. \label{stxcs}
\ea
Therefore, the self-similar vacuum solution is nothing but the zero-mass static solutions discussed in Appendix \ref{sec-static}.  The nature of the singularity at $r=0$ depends on the value of the parameters $\gamma$ and $\kappa$.  

The static $(r,t)$ coordinates are best suited for determining the causal structure of the spacetimes. Let us bear in mind the discussion on the restrictions on these parameters from Appendix \ref{sec-static} --- we will continue to impose these restrictions for the remainder of this article as they are physically well motivated.   If $V_{\mathrm{eff}} \neq 0$ and $4 / ( 8 - \gamma ) < \kappa < 2/(4-\gamma)$,  then there is a timelike singularity at $r=0$.    Demanding only null singularities, one has $\kappa > 2/(4-\gamma)$.\footnote{Actually,   the value $\kappa = 2/ (4-\gamma)$ with  is $V_{\mathrm{eff}} \neq 0$ is a special case, for which the curvature is zero everywhere --- the strong coupling surface (where $\phi \to \infty$) is timelike --- it is analogous to the origin of the spherically symmetric coordinate system. in higher dimensions.  We will study this special case later.}

\begin{figure}
\centering
\begin{subfigure}{.50\textwidth}
  \centering
  \includegraphics[height=2in]{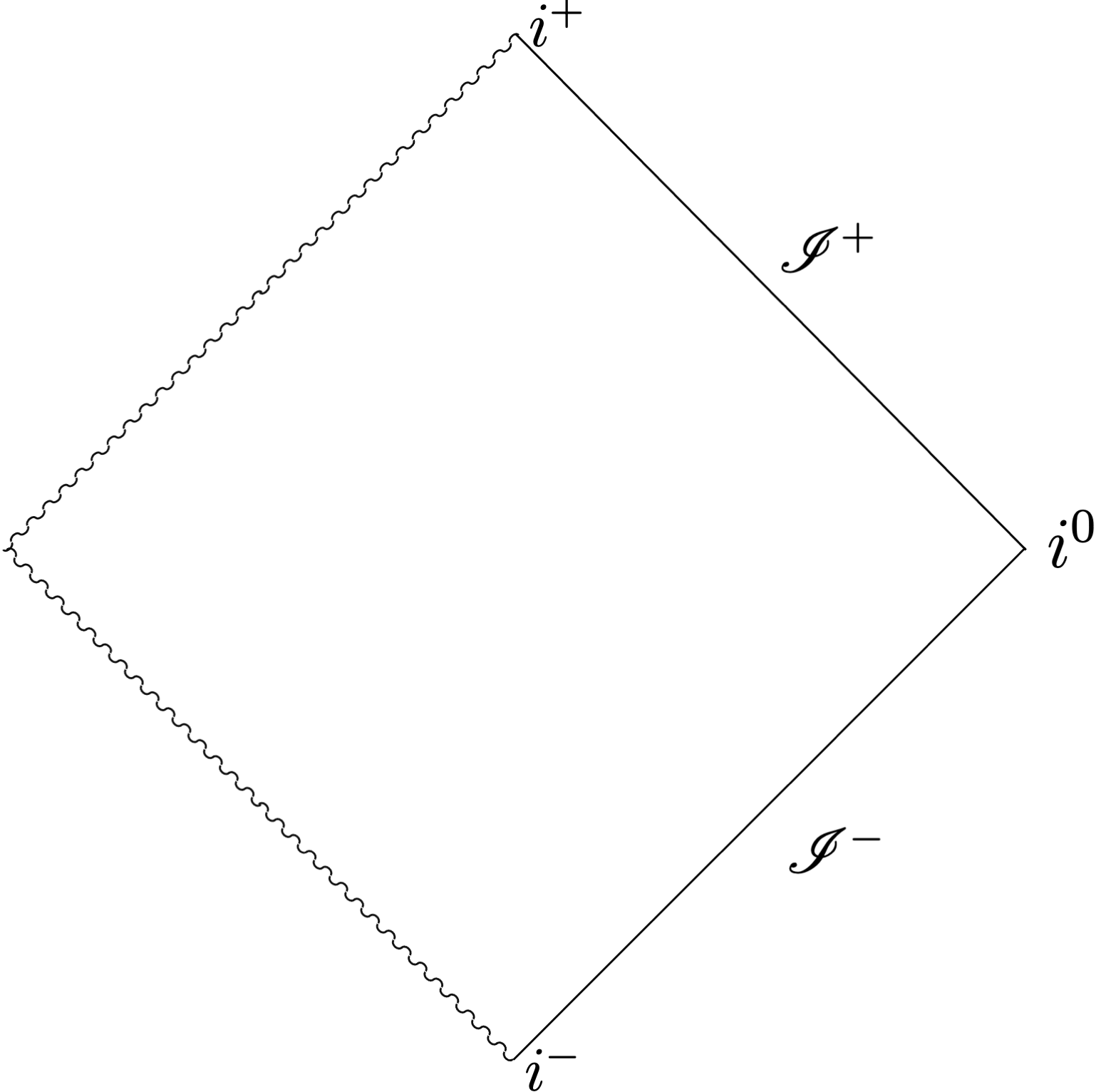}
  \caption{}
  \label{fig:cpa}
\end{subfigure}%
\begin{subfigure}{.25\textwidth}
  \centering
  \includegraphics[height=2in]{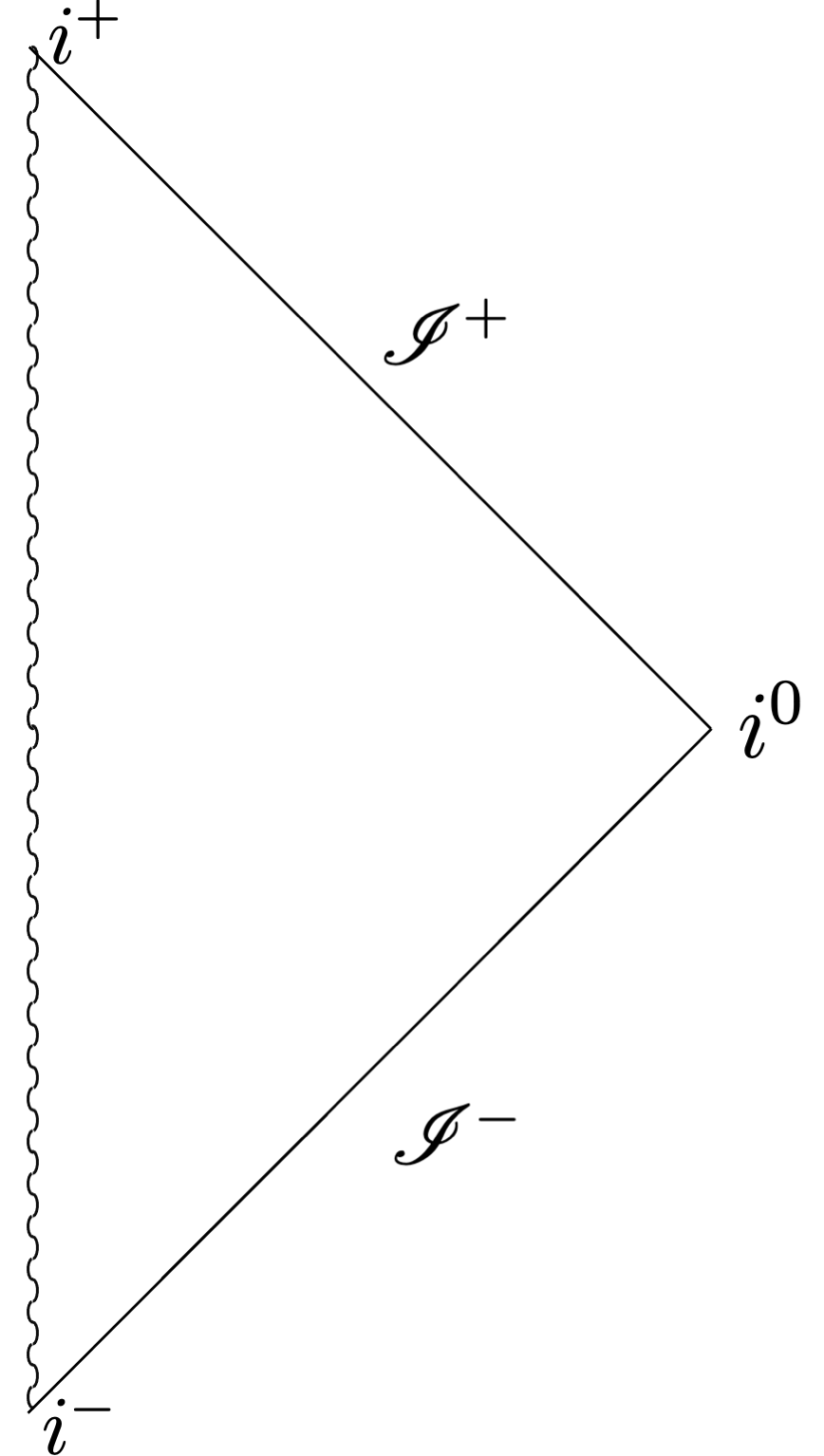}
  \caption{}
  \label{fig:cpb}
\end{subfigure}%
\begin{subfigure}{.25\textwidth}
  \centering
  \includegraphics[height=2in]{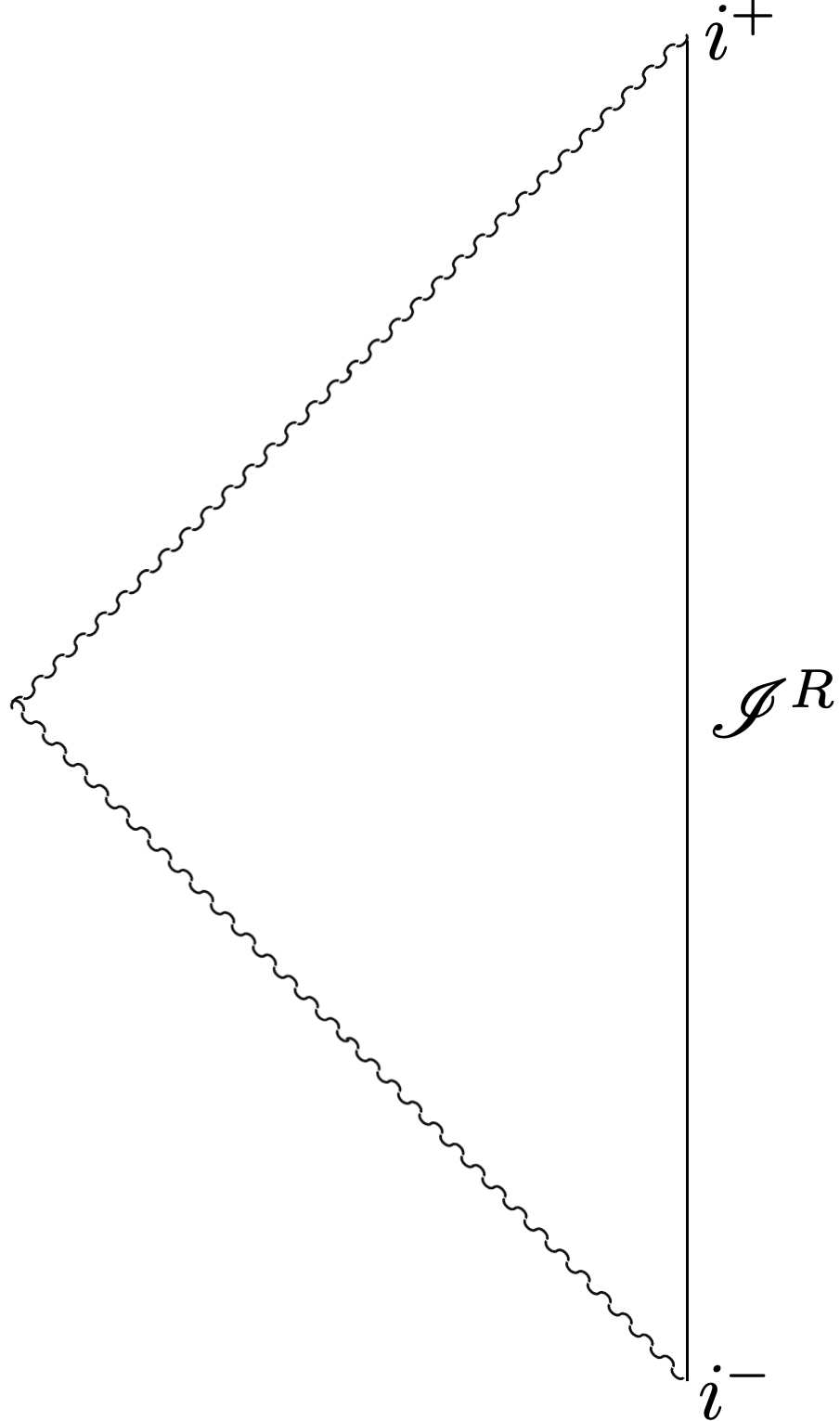}
  \caption{}
  \label{fig:cpc}
  \end{subfigure}\caption{Possibilities for the Carter-Penrose diagrams for the self-similar spacetimes in this section:  (a) and (b) correspond to an asymptotic causal structure similar to Minkowski, while (c) corresponds to an AdS-like asymptotic causal structure.  The wavy lines represent curvature singularities or regions of strong coupling, in case the curvature is zero everywhere.   }
\label{fig:cp}
\end{figure}

It is worth pointing out that for $\kappa > \flatfrac{2}{(4-\gamma)}$,  $v=0$ corresponds to a curvature singularity and hence,  it is not possible to continue the $v$ coordinate to the region $v<0$.   Furthermore,   the map \eqref{stxcs}  between the static coordinates and the self-similar coordinates obscures an easy intuition about the self-similar coordinates,  which is rather straightforward in the static coordinate system.  In spite of this issue,  we note a useful fact about the $(\rho, v)$ coordinates for all the systems: $x\to \infty$ with finite non-zero $v$ describes the asymptotic region $r\to \infty$, where the curvature is vanishingly small for our choice of range of parameters.

For the stringy coupling, $\gamma = 4$, we have,
\baa
F(x)  =  \frac{\kappa  V_{\mathrm{eff}} }{4 (\kappa -1)} \exp(-\frac{2  x}{\kappa }  ) - 2 \kappa \label{fxg4},
\ea
and,
\baa
\Phi(x) = -x . \label{phixg4}
\ea
This is equivalent to the $r_h \to -\infty$ limit of the metric function \eqref{frg4}, i.e.,  the line element,
\baa
\dd s^2 =  \frac{\kappa  V_{\mathrm{eff}} }{4 (\kappa -1)} \exp(-\frac{2  r}{\kappa }  ) \dd \bar{v}^2 + 2 \, \dd{v} \dd{r}.  \label{mswstr}
\ea
In this case,  for $V_{\mathrm{eff}} \neq 0$,  the curvature grows unboundedly as $r\to -\infty$,  which describes a timelike ``singularity'' (see Figure \ref{fig:cpc}).  However,  as mentioned in Appendix \ref{sec-static},  this is not a true singularity \cite{Wald:1984rg}, since it takes an infinite amount of affine time for a null geodesic to reach this singularity.   For $V_{\mathrm{eff}} = 0$,  the curvature is zero everywhere. 
The coordinate transformation that brings this metric to the form described by \eqref{fxg4} is given by,
\baa
r &= \frac{\rho}{v} + \kappa \log v,  \\
\bar{v} &= \frac12 v^2. \label{ctrag4}
\ea

Finally, for the Liouville system, the self-similar solution is given by,
\baa
\Sigma (x) = \log x,  \label{liosx}
\ea
(here, we use  notation analogous to \eqref{cssphi},  i.e,  $\sigma( r , v) = \Sigma (x) - \kappa \log v$) and,
\baa
F(x) = \frac{V_{\mathrm{eff}} \kappa }{2+ \kappa } x^{ 2 + \frac{2}{\kappa } } + 2 \kappa x. \label{lioxf}
\ea
The coordinate transformation from the $r_h \to 0$ static metric (see eq.  \eqref{liofr}) is given by,
\baa
r &= \frac{\rho}{v^{\kappa +1}},  \\
\bar v &= \frac{1}{\kappa+2} v^{\kappa +2}.  \label{liocot}
\ea
Note that we have taken $\kappa < -2$.  See the discussion in the paragraph before eq.  \eqref{fxg4} and Appendix \ref{sec-static}.

Let us also emphasise here that we need a large value of $\sigma_0$ (the coefficient of the topological term in \eqref{lioac}).  Since the self-similar solution is the $r_h \to 0$ limit of the static solution,  the singularity of the geometry,  at $r = 0$,   is null in nature.  However,  strictly speaking if we want $\sigma_0 + \sigma$ to be positive at finite $\sigma_0$,   $r=0$ is not part of the spacetime and the spacetime terminates on the timelike surface $r = e^{-\sigma_0}$.   To ensure that the curvature singularity coincides with this surface,  we always take the limit $\sigma_0 \to \infty$.  Since $\sigma_0$ does not affect the dynamics,  we will only need to determine the curvature singularity.  In the remainder of this article,  we shall be using this limit. See also Appendix \ref{sec-static}.

We therefore see why additional matter truly matters ---  all the vacuum self-similar solutions are just zero mass static solutions. To make things more interesting we add to our systems matter contents --- mainly scalar fields,  consistent with the assumption of homothety.  We will see that addition of matter fields makes the dynamics quite rich and interesting.

At the risk of repetition,  it is worth emphasising here that the scenario being considered in this subsection is  different from gravity coupled to matter in higher dimensions --- the gravity-dilaton system here is more similar to pure gravity in higher dimensions. What is truly interesting here is the fact that a certain non-trivial static solution is also a non-trivial self-similar solution. This feature is special to two-dimensional dilaton gravity and can happen only trivially in higher dimensions --- i.e.,  the Minkowski spacetime.

Let us also point out that under the assumption of continuous self-similarity,  the only possible static solution that could have been self-similar is the zero mass solution.  If  there were a mass present,  it would have implied the presence of a length scale in the geometry, which is inconsistent with CSS.

\subsection{Autonomous Equations for the Stringy System with Matter}\label{subsec-striau}

Let us consider adding matter to the stringy system first. We consider a single scalar field $\psi$.  If we assume that this scalar has the same sort of homothety property \eqref{ldevkap} as the dilaton,    i.e.,
\baa
\pounds_\xi \psi = - \lambda, \label{scalho}
\ea
then the first obvious candidate --- a minimally coupled scalar with zero mass --- is actually not consistent with homothety. This can be readily fixed by considering a non-minimal dilaton coupling to the scalar --- we simply add the common pre-factor $\exp(-2\phi)$.  For good measure,  we could also add a potential term for the scalar so that the combined action reads,
\baa
S =  \frac{1}{16 \pi G_2} \int \dd[2]{x} \, \sqrt{-g} e^{-2\phi} \bqty{ R + \gamma (\nabla \phi)^2 + V_{\mathrm{eff} } e^{2\phi/\kappa} - (\nabla \psi)^2 - V_\psi e^{2\psi/\lambda}   }.  \label{phimatac}
 \ea
One could think of trying to incorporate scalar matter without the dilaton coupling --- it appears to be quite complicated.  One would not have a simple relation of the form \eqref{scalho} and the homothety properties of the scalar might be dependent on other fields --- which is not a very attractive scenario.

From the point of view of gravity in higher spacetime dimensions, however,   this particular form of dilaton coupling with matter is the appropriate one.  If one starts with a $(d+2)$-dimensional gravitational theory with a minimally coupled scalar $\psi$, (with the potential term as well), then after compactifying down to two dimensions,  the scalar kinetic term and the Ricci scalar will have the same pre-factor. This is what happens here.
 
The $\phi$,  $g^{\mu \nu}$ and $\psi$ equations of motion are given by,
 \baa
R - \gamma ( \nabla \phi )^2 + \gamma \nabla^2 \phi+ V_{\mathrm{eff} }  \pqty{1-\frac{1}{\kappa }} \exp(  \frac{2  \phi}{\kappa} ) - (\nabla \psi)^2 - V_\psi e^{2\psi/\lambda}  =0, \label{eom28}
\ea
\baa
2  \nabla_\mu \nabla_\nu \phi - 2 g_{\mu \nu}  \nabla^2 \phi  & + \frac{8-\gamma}{2} g_{\mu \nu}  (\nabla \phi)^2 + (\gamma -4) \nabla_\mu \phi \nabla_\nu \phi  - \frac12 g_{\mu \nu} e^{  2  \phi/\kappa }  V_{\mathrm{eff} } \\
& - \pqty{ \nabla_\mu \psi \nabla_\nu \psi - \frac12 g_{\mu \nu} (\nabla \psi)^2  } + \frac12 V_\psi e^{2\psi/\lambda} g_{\mu \nu} = 0, \label{eom29}
\ea
and,
\baa
4 \nabla_\mu \phi \nabla^\mu \psi - 2 \nabla^2 \psi  + \frac{2}{\lambda} V_\psi e^{2\psi/\lambda} = 0, \label{eom30}
\ea
respectively.  

We use the ans\"{a}tze  \eqref{cssan},  \eqref{cssphi} and the one that follows from \eqref{scalho}, i.e,
\baa
\psi (\rho, v) = \Psi ( x ) - \lambda \log v.  \label{csspsi}
\ea
Inserting these in the equations of motion yields ordinary differential equations in the independent variable $x$.

The set of equations of motion are not all independent.  For example,   \eqref{eom28} follows from the \eqref{eom30} and the $(rr)$ and $(rv)$ components of \eqref{eom29}.  
The assumption of self-similarity imposes a stronger constraint than the static case,  and the metric function $F(x)$ is \emph{uniquely} (i.e., not even up to a constant) determined by $\Phi (x),  \Psi(x)$ and their first derivatives. The only dynamical equations to solve are given by a pair of autonomous second order ordinary differential equations,
\baa
\Phi ''(x) =  \frac{1}{2} \left( (4-\gamma) \Phi '(x)^2 + \Psi '(x)^2 \right),  \label{phidp1}
\ea
and,
\baa
\Psi'' (x) = \frac{\mathcal{N}}{\lambda  \left(\kappa  \left(V e^{\frac{2 \Phi (x)}{\kappa }}-V_\psi  e^{\frac{2 \Psi (x)}{\lambda }}\right)+2 \left(\lambda ^2-\kappa  ((\gamma -8) \kappa +4)\right) \Phi '(x)\right) }, \label{psidp1}
\ea
where the numerator $\mathcal{N}$ is given by,
\baa
\mathcal{N} &= \Psi '(x)^2 \left(\lambda ^2 V_{\mathrm{eff} } e^{\frac{2 \Phi (x)}{\kappa }}-V_\psi \left(\kappa +\lambda ^2\right) e^{\frac{2 \Psi (x)}{\lambda }}+\lambda  \left(\lambda ^2-\kappa  ((\gamma -6) \kappa +4)\right) \Psi '(x)\right) \\
&\quad +\lambda  \Phi '(x) \Psi '(x) \left(- V_{\mathrm{eff} } ((\gamma -6) \kappa +2) e^{\frac{2 \Phi (x)}{\kappa }}+V_\psi ((\gamma -6) \kappa +4) e^{\frac{2 \Psi (x)}{\lambda }}+6 \kappa  \lambda  \Psi '(x)\right) \\
&\quad +\Phi '(x)^2 \left(\lambda  \left((\gamma -6) \kappa  ((\gamma -8) \kappa +4)-\gamma  \lambda ^2\right) \Psi '(x)-V_\psi ((\gamma -8) \kappa +4) e^{\frac{2 \Psi (x)}{\lambda }}\right) \\
&\quad +2 \lambda ^2 ((\gamma -8) \kappa +4) \Phi '(x)^3.  \label{psidpn}
\ea

We wrote the above pair of equations to illustrate the autonomous  nature of the differential equations. It would perhaps be helpful to write down the equations in a simpler form.  We will also be able to explain easily how we derived these equations.

The equation \eqref{psidp1} is actually nothing but the $(\rho \rho)$ component of the metric equation \eqref{eom29}. Inserting \eqref{eom29} in the $(v \rho)$ component of the metric equation, we can obtain a simple expression for $F'(x)$ in terms of  $\Phi (x)$,  $\Psi (x)$, their derivatives and $F(x)$:
\baa
F'(x) = \frac{V_{\psi } e^{\frac{2 \Psi (x)}{\lambda }}-V_{\text{eff}} e^{\frac{2 \Phi (x)}{\kappa }}+\gamma  F(x) \Phi '(x)^2-F(x) \Psi '(x)^2-8 \kappa  \Phi '(x)}{2 \Phi '(x)}.  \label{fp1xf}
\ea
We can now insert \eqref{fp1xf} and \eqref{phidp1} in the last remaining metric equation,  i.e., the $(vv)$ component.  Solving the resulting equation for $F(x)$ yields $F(x)$ of the promised form:

\baa
F(x) = \frac{\kappa  \left(  V_{\psi } e^{\frac{2 \Psi (x)}{\lambda }}-V_{\text{eff}}  e^{\frac{2 \Phi (x)}{\kappa }}\right)+2 \left(\kappa  ((\gamma -8) \kappa +4)-\lambda ^2\right) \Phi '(x)}{((\gamma -8) \kappa +4) \Phi '(x)^2+\kappa  \Psi '(x)^2-2 \lambda  \Phi '(x) \Psi '(x)}. \label{fxsolphi}
\ea

If we now look at the $\psi$ equation of motion \eqref{eom30} and solve for $\Psi''(x)$,  we obtain,
\baa
\Psi''(x) = \frac{V_{\psi } e^{\frac{2 \Psi (x)}{\lambda }}-\lambda  \Psi '(x) \left(F'(x)-2 F(x) \Phi '(x)+2 \kappa \right)-2 \lambda^2  \Phi '(x)}{\lambda  F(x)}.  \label{psidp2}
\ea
Inserting \eqref{fp1xf} and \eqref{fxsolphi} in \eqref{psidp2} yields the rather complicated expression that is \eqref{psidp1}.

Complicated as the equation \eqref{psidp1} may be,  we can make some non-trivial statements by virtue of the simplicity of the equation \eqref{phidp1}.   Since the equations \eqref{phidp1}  and \eqref{psidp1} are a pair of second-order autonomous equations,  our first instinct would be to convert each second order equation into a pair of first order equations by defining,
\baa
\Phi'(x) &= P_\Phi (x),  \\ \quad \Psi'(x) &= P_\Psi (x).  \label{foode}
\ea
The two equations \eqref{phidp1}  and \eqref{psidp1} can now be written as $P'_\Phi (x) = \cdots$ and $P'_\Psi (x) = \cdots$, where the same expressions as the right-hand-sides of \eqref{phidp1}  and \eqref{psidp1} appear in the ellipses, with the replacement  \eqref{foode}.

Given an autonomous system such as the above,  we would like to find the fixed points of the system and perform a linear stability analysis around the fixed point. In case of the system in question,  such a fixed point represents an entire spacetime.

Looking at eqs.  \eqref{phidp1},  \eqref{psidp1} and \eqref{foode},  we see immediately that the fixed points are characterised by $P_\Phi = 0 = P_\Psi$. Since the phase space of the autonomous system is four-dimensional, we actually have a co-dimension two surface,  i.e. a two-dimensional plane of non-isolated fixed points.  If we set $V_\psi = 0$, then only $\Psi'(x)$ appears in the differential equations \eqref{phidp1},  \eqref{psidp1}  and not $\Psi(x)$. Therefore, there is an effective reduction of phase space and the phase space now becomes three-dimensional --- i.e.,  the second of \eqref{foode} is just a definition and no longer an equation governing autonomous evolution.  In this case, too,  the fixed points are co-dimension two,  (i.e.,  a line of fixed points) given by $P_\Phi = 0 = P_\Psi$.

Let us now come to the issue of linear stability of these fixed points.  For the four-dimensional phase space (when $V_\psi \neq 0$),  the non-isolated fixed points lie on a plane. We therefore immediately see that two of the eigenvalues about each fixed point would be zero. These correspond to constant values $\Phi(x) = \Phi_0$ and $\Psi (x) = \Psi_0$.   What would be interesting is the behaviour along the remaining two dimensional subspace even after we fix $\Phi, \Psi$ to some constant values.  It is quite remarkable that the remaining eigenvalues are zero as well.  To see this, note that all the terms on the RHS of \eqref{phidp1} and the numerator of \eqref{psidp1} are at least of quadratic order in $P_\Phi$ or $P_\Psi$. Therefore, taking a derivative with respect to either $P_\Phi$ or $P_\Psi$ and then setting  $P_\Phi = 0 = P_\Psi$ yields zero.  Therefore,  the eigenvalues are zero.

The results in the previous paragraph are in sharp contrast with self-similar collapse models studied in higher dimensional spacetimes.  In such studies, there were typically more than one fixed point of the autonomous system of differential equations.   One of the fixed points was associated with the flat spacetime.   One of the fixed points usually corresponded to a black hole spacetime.  Previous studies also found a saddle point which governed the critical behaviour and was responsible for Choptuik-like scaling. In our case,  however, linearisation yields no significant insight into the dynamics --- clearly,  we should explore different options to understand the solutions to the differential equations \eqref{phidp1} and \eqref{psidp1}.

Taken as a statement on the differential equations \eqref{phidp1} and \eqref{psidp1} the comments made above the linearisation are true.  However,  when we think a little more carefully and take the physics into account, we will conclude that the particular linearisation we discussed is itself not a valid step.  To see this, we can go back to the  equations of motion and ask whether the constant solutions,   viz.,  $\Phi (x) = \Phi_0$ and $\Psi (x) = \Psi_0$ meet the equations of motion.  It turns out that the constants are valid solutions only if the constraint 
\baa
V_{\psi } \exp(\frac{2 \Psi_0 }{\lambda })-V_{\text{eff}}  \exp(\frac{2 \Phi_0}{\kappa }) =0 \label{potcon1}
\ea
is met.   When we impose this constraint in conjunction with $P_\Phi  = 0 = P_\Psi$,  we find that the right hand side of \eqref{psidp1} becomes of the form $0/0$.  We will therefore adopt new strategies to attack the problem in the following sections.  Before we do that, however,  let us obtain the requisite differential equations for the Liouville system coupled to two different classes of matter.

\subsection{Liouville System and a Minimally Coupled Scalar}\label{subsec-limau}

In the previous subsection we discussed the difficulty of introducing a minimally coupled scalar field into the picture while respecting the homothety symmetry.  We now consider the Liouville system earlier discussed in \S\ref{subsec-lioset}. Thanks to the nature of the couplings in this case,  it becomes possible to introduce a minimally coupled scalar field $\psi$ (with perhaps an additional appropriate potential) that satisfies the homothety property \eqref{scalho}. The action is given by,
\baa
S = \frac{1}{16\pi G_2} \int \dd[2]{x} \sqrt{-g} \pqty{ (\sigma_0 + \sigma) R - ( \nabla \sigma)^2 + V_{\mathrm{eff}} e^{2\sigma/\kappa}  - (\nabla \psi)^2 -  V_\psi e^{2\psi/\lambda}  }.  \label{sigmac}
\ea

The equations of motion obtained by varying $\sigma$,  $g^{\mu \nu}$ and $\psi$ are
\baa
R + 2 \nabla^2 \sigma + \frac{2}{\kappa} V_{\mathrm{eff}} e^{2\sigma/\kappa}  &= 0, \label{simeom}
\ea
\baa
&g_{\mu \nu} \nabla^2 \sigma - \nabla_\mu \nabla_\nu \sigma - \pqty{  \nabla_\mu \sigma \nabla_\nu \sigma - \frac12 g_{\mu \nu} (\nabla \sigma)^2 } - \frac12 g_{\mu \nu} V_{\mathrm{eff}} e^{2\sigma/\kappa}  \\
& - \pqty{ \nabla_\mu \psi \nabla_\nu \psi - \frac12 g_{\mu \nu} (\nabla \psi)^2  } + \frac12 V_\psi e^{2\psi/\lambda} g_{\mu \nu} = 0,  \label{simeom2}
\ea
and,
\baa
\nabla^2 \psi - \frac{1}{\lambda} V_\psi e^{2 \psi / \lambda} &= 0,  \label{simeom3}
\ea
respectively.

We use the ans\"{a}tze \eqref{cssan} and \eqref{csspsi} for the metric and the scalar $\psi$ and use the analogous ansatz for $\sigma$,
\baa
\sigma (\rho , v) =  \Sigma (x) - \kappa \log v.   \label{csssig}
\ea
Since we have described the process in some detail in the previous subsection,  we will be brief here.  As before,  we obtain second order autonomous differential equations for $\Sigma (x)$  and $\Psi (x)$, which are given by,
\baa
\Sigma ''(x) =  -\Sigma '(x)^2-\Psi '(x)^2,  \label{scssm1}
\ea
and
\baa
\Psi ''(x) = -\frac{\lambda  F'(x) \Psi '(x)-V_{\psi } e^{\frac{2 \Psi (x)}{\lambda }}}{\lambda  F(x)}, \label{scssm2}
\ea
respectively, where,
\baa
F'(x) = \frac{V_{\text{eff}} e^{\frac{2 \Sigma (x)}{\kappa }}+F(x) \Sigma '(x)^2+F(x) \Psi '(x)^2-V_{\psi } e^{\frac{2 \Psi (x)}{\lambda }}}{\Sigma '(x)}, \label{fpxsig1}
\ea
and,
\baa
F(x)= \frac{\kappa  V_{\text{eff}} e^{\frac{2 \Sigma (x)}{\kappa }}-\kappa  V_{\psi } e^{\frac{2 \Psi (x)}{\lambda }}+2 \left(\kappa  (\kappa +2)+\lambda ^2\right) \Sigma '(x)}{(\kappa +2) \Sigma '(x)^2-\kappa  \Psi '(x)^2+2 \lambda  \Sigma '(x) \Psi '(x)}.   \label{fxsig11}
\ea
Similar comments as those after \eqref{psidp2} apply in this case too.  These appear to be quite generic features (e.g., limited utility of the linear stability analysis around a fixed point) of the systems we have considered so far.

Let us point out an interesting fact here.  The Ricci scalar \eqref{ricUR} is determined in this case only by the potential functions,
\baa
 F''(x) = \frac{2 (\kappa +1) V_{\text{eff}} e^{\frac{2 \Sigma (x)}{\kappa }}}{\kappa }-2 V_{\psi } e^{\frac{2 \Psi (x)}{\lambda }}.  \label{sigric}
\ea
It can be checked that such is not the case for the stringy system --- derivatives of $\Phi,  \Psi$ terms do contribute to the scalar curvature. 
\subsection{Liouville System and a Semi-classical Model}\label{subsec-liscau}
As mentioned in the introduction,  two-dimensional gravity is a rich playground for dealing with tractable aspects of quantum gravity.  Models in which quantum corrections are under calculational control typically involve some sort of semi-classical approximation.  One realisation of such a scenario is given by adding $N$ free massless scalar fields, 
\baa
S_{\mathrm{M} } = - \frac12 \int \dd[2]{x} \sqrt{-g} \sum_i (\nabla \psi_i)^2.  \label{scaln}
\ea
and studying the back-reaction of these scalars in the large $N$ limit.  More precisely,  one demands that $N\to \infty$ and the two-dimensional Newton constant $G_2 \to 0$ in such a way that $G_2 N =$ finite.   {Such a semi-classical set-up would remind the reader of the Callan-Giddings-Harvey-Strominger (CGHS) model \cite{Callan:1992rs} in 2D gravity.  The scenario here is slightly different.  Had we added this matter content to the stringy model (the pure-gravity sector of \eqref{phimatac}) with $\gamma =4$ and $\kappa \to \infty$,   this would have been completely equivalent to the CGHS model.  As mentioned previously,  with minimally coupled matter in the stringy model,  the homothety transformation property would  be very complicated,  if it exists. }

A particular aspect of the quantum-corrected system, namely the trace anomaly,  is captured by a simple classical action involving a scalar $\chi$ that also couples to the Ricci scalar \cite{Almheiri:2014cka} (see also \cite{Moitra:2019xoj}),
\baa
S_{\mathrm{M} } = - \frac{N}{24 \pi} \int \dd[2]{x} \sqrt{-g} ( \chi R + (\nabla \chi)^2 ).  \label{chiac}
\ea
Addition of this matter action to the Liouville system (i.e.,  eq.  \eqref{sigmac} without the $\psi$) is consistent with the homothety properties discussed so far.   The $\sigma$ equation of motion is the same as before,  i.e.,  eq.  \eqref{simeom}.  The $g^{\mu \nu}$ and the $\chi$ equations of motion are given by,
\baa
{} g_{\mu \nu} \nabla^2 (\sigma - \zeta \chi) - \nabla_\mu \nabla_\nu (\sigma - \zeta \chi) & - \pqty{  \nabla_\mu \sigma \nabla_\nu \sigma   - \frac12 g_{\mu \nu} (\nabla \sigma)^2 } \\
  - \frac12 g_{\mu \nu} V_{\mathrm{eff}} e^{2\sigma/\kappa}  &- \zeta \pqty{ \nabla_\mu \chi \nabla_\nu \chi - \frac12 g_{\mu \nu} (\nabla \chi)^2  }  = 0,  \label{meomchi}
\ea
and,
\baa
R - 2 \nabla^2 \chi = 0,  \label{cheomch}
\ea
respectively. Note that we have defined the quantity,
\baa
\zeta = \frac23 GN, \label{zetadef}
\ea
which is the effective quantity controlling the semi-classical dynamics.

Taking, as before,  $\pounds_\xi \chi = - \lambda$,  we get,
\baa
\chi ( \rho, v) = X (x) - \lambda \log v, \label{chirhov}
\ea
using which we can analogously obtain the two second order differential equations in $\Sigma$ and $X$.

\baa
\Sigma ''(x) -  \zeta  X''(x) = -\zeta  X'(x)^2-\Sigma '(x)^2,  \label{sigmadoubleprime}
\ea
\baa
X''(x) = \frac{-F''(x)-2 F'(x) X'(x)}{2 F(x)},  \label{eksdoubleprime}
\ea
where the metric function $F(x)$ and its derivative are given by,
\baa
F(x) = \frac{V_{\text{eff}} (\kappa -\zeta  \lambda ) e^{\frac{2 \Sigma (x)}{\kappa }}+2 (\zeta  (\lambda -2) \lambda +\kappa  (\kappa +2)) \left(\Sigma '(x)-\zeta  X'(x)\right)}{-2 \zeta  (\kappa -\lambda +2) \Sigma '(x) X'(x)+(\zeta  \lambda +\kappa +2) \Sigma '(x)^2-\zeta  (\zeta  (\lambda -2)+\kappa ) X'(x)^2},  \label{Fparxfull}
\ea
and
\baa
F'(x) =  \frac{V_{\text{eff}} e^{\frac{2 \Sigma (x)}{\kappa }}+F(x) \left(\zeta  X'(x)^2+\Sigma '(x)^2\right)}{\Sigma '(x)-\zeta  X'(x)} \label{Fparpxfull}
\ea
respectively.  The second derivative of $F(x)$,  which determines the Ricci scalar,  is determined only by the field $\Sigma(x)$, like in the previous model,
\baa
F''(x) = \frac{2 (\kappa +1) V_{\text{eff}} e^{\frac{2 \Sigma (x)}{\kappa }}}{(\zeta +1) \kappa }.  \label{Fparppxfull}
\ea

It is worth contrasting the result here with a previous work, \cite{Chiba:1997ex} which had ruled out the possibility of continuously self-similar solutions in a certain model of two-dimensional semi-classical gravity.   {The scenario in \cite{Chiba:1997ex} is rather different from the one explored in this subsection: the authors examined the semi-classical equation of motion in a particular model and concluded that the two sides of the equation did not transform covariantly under the homothety transformation.  Our semi-classical model is different from theirs and,  as is obvious above, consistent with continuous self-similarity.}

\section{Interlude: Some Simple Dynamical Systems}\label{sec-interlude}

The results of the previous section suggest that linear stability analysis is perhaps not the best approach in dealing with the systems under consideration.  Quite interestingly,  we will be able to say something about singularity formation in these models precisely because of the reason linear stability analysis fails.

\begin{figure}
\centering
\begin{subfigure}{.33\textwidth}
  \centering
  \includegraphics[width=0.9\linewidth]{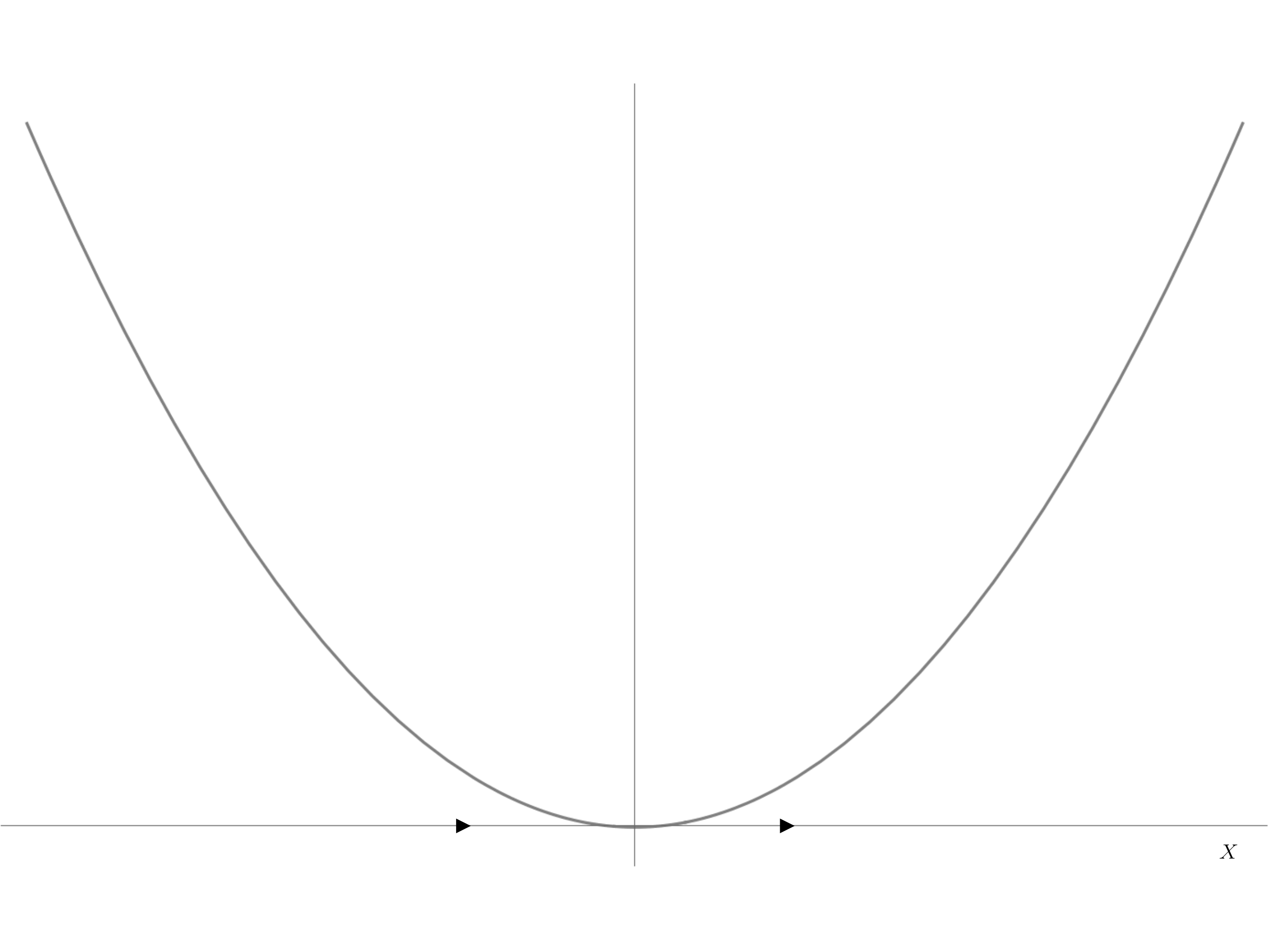}
  \caption{}
  \label{fig:dynsysa}
\end{subfigure}%
\begin{subfigure}{.33\textwidth}
  \centering
  \includegraphics[width=0.9\linewidth]{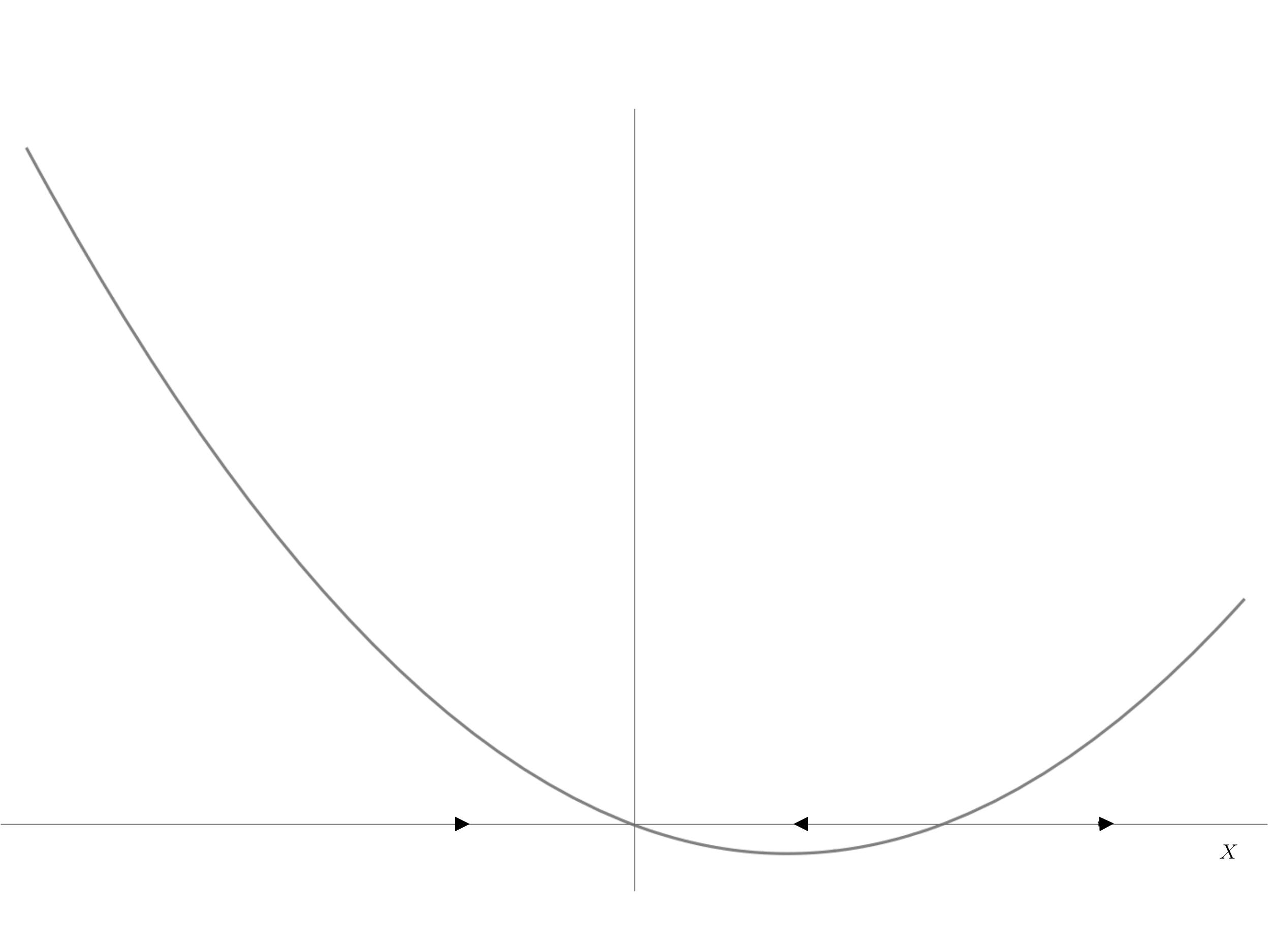}
  \caption{}
  \label{fig:dynsysb}
\end{subfigure}%
\begin{subfigure}{.33\textwidth}
  \centering
  \includegraphics[width=0.9\linewidth]{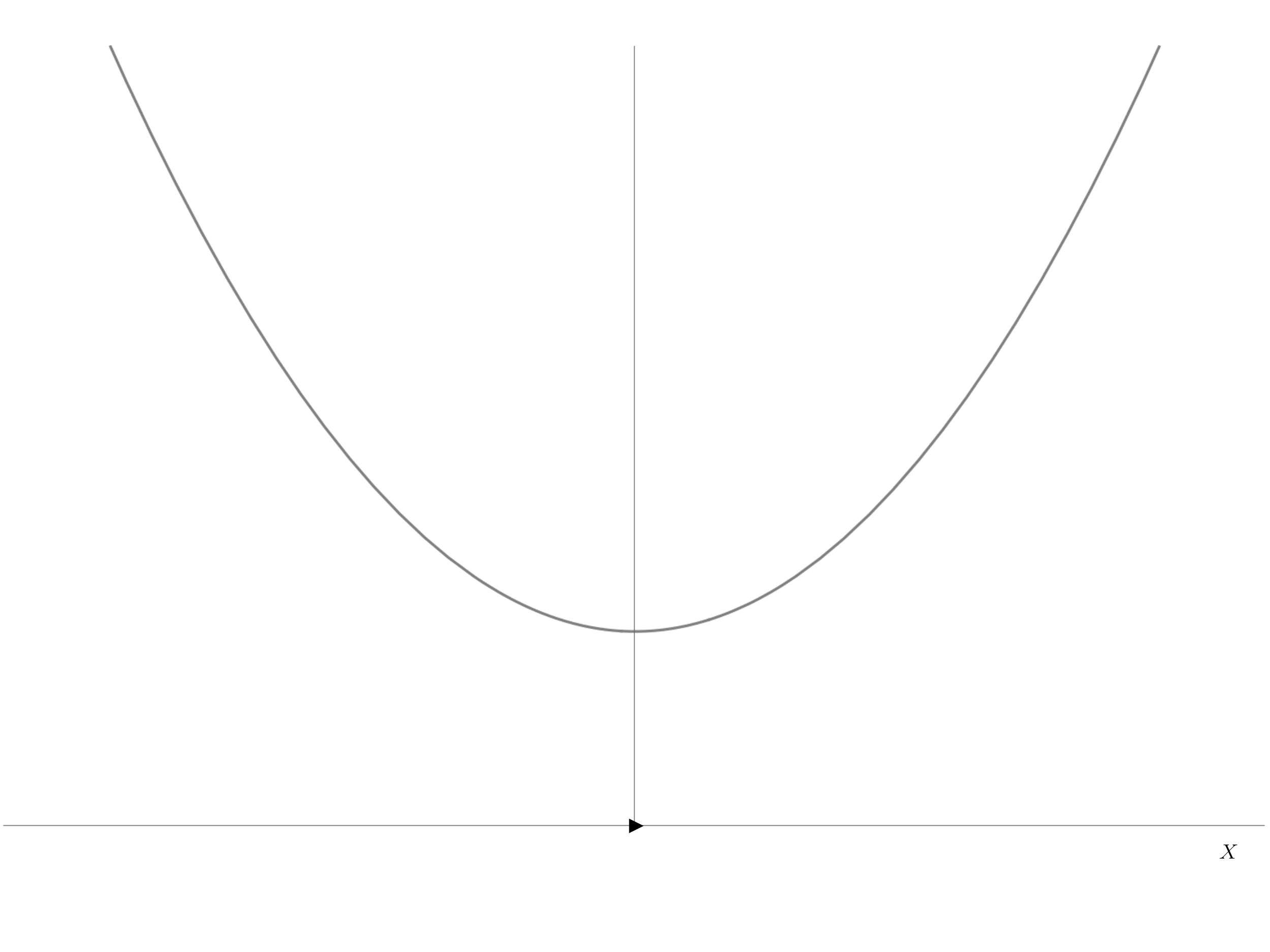}
  \caption{}
  \label{fig:dynsysc}
\end{subfigure}
\caption{The phase plots for the dynamical systems described by eqs.  \eqref{sys1}, \eqref{sys2} and \eqref{sys3} respectively}
\label{fig:dynsys}
\end{figure}

Let us consider first the one-dimensional dynamical system governed by the equation,
\baa
\dot X =X^2, \label{sys1}
\ea
where the overdot ($\dot{}$) indicates a time-derivative.  The equation  is suitably non-dimensionalised. The phase plot for the system is shown in Figure \ref{fig:dynsysa}.

This system has a fixed point at $X=0$.  However,  linear stability analysis does not yield any useful information about the dynamics as $ \eval{\dv*{(X^2)}{X}}_{X=0} = 0$. This fixed point is clearly half-stable.  This particular dynamical system has one remarkable feature,  though, which goes by the name of \emph{finite-time blow-up} \cite{strogatz2000nonlinear}.   In quantum field theory,  an analogous phenomenon is a divergent coupling constant at a finite energy scale \cite{Weinberg:1996kr}.   As we argue now,  this feature is intimately related to spacetime singularities in the two-dimensional theory of gravity.

If we have $X = 0$ at any time, then clearly, we shall have $X=0$ for all times,  as follows from \eqref{sys1}.  We ask what happens when we consider slight perturbations from the equilibrium solution.  Let us solve the equation with the initial condition $X(0) = \epsilon$ with $|\epsilon | \ll 1$.
The solution is, of course,
\baa
X(t) =  \frac{1}{\frac{1}{\epsilon}-t}. \label{xt1bept}
\ea
If $\epsilon >0$, then $X(t) \to \infty$ as $t \to \epsilon^{-1}$ --- i.e.,  an infinite value of $X$ is reached in a \emph{finite} time.  If we had assumed $\epsilon < 0$, then we would have concluded that there was a blow-up $X \to -\infty$ at a finite time in the \emph{past}.

The flow from one linear fixed point to another usually takes an infinite amount of time.  However,  if the dynamics of the system is dominated by non-linear terms, then the occurrence of finite-time blow-up is quite generic.  Let us consider the system (see Figure \ref{fig:dynsysb}) ,
\baa
\dot{X} = - X + X^2,  \label{sys2}
\ea
In this case, there are linear fixed points at $X=0$ and $X=1$.  The fixed points at $X=0$ and $X=1$ are stable and unstable respectively.  If at $t=0$, $0<X<1$, then $X(t\to -\infty) = 0$ and $X(t \to -\infty) = 1$.  However, even though both are linear fixed points, this system shows finite blow-up in the future if the initial condition satisfies $X(0) >1$,  since the dynamics is now dominated by the $X^2$ term in \eqref{sys2}. If $X(0) <0$,  then the system shows finite-time blow-up in the past due to the eventual dominance of $X^2$.

{We can understand the phenomenon of finite-time blow-up more physically.  We can imagine a particle whose dynamics is governed by an equation  of the form $\dot{X} = X^\alpha$. For $\alpha > 0$,  a particle released from $X = \epsilon$ continues to travel with increasing speed to the right. For $0<\alpha<1$,  the increase in speed is slow, at most polynomial in time.  When $\alpha = 1$,  the behaviour is exponential in time.  When $\alpha >1$,  the process is even faster --- it shows a runaway behaviour and the particle reaches any point in finite time.  In most physical systems, such runaway behaviour is typically opposed by higher order effects. }

In the context of the present article,  the relevance of the interlude is hopefully clear.  We encounter precisely the same form of equation as \eqref{sys1} for the self-similar systems in the absence of matter,  see e.g.,  eqs.  \eqref{phidp1} and \eqref{scssm1}.  In this section, the time variable $t$ plays the role of the  self-similar variable $x$ and $X$ is equivalent to the derivative of the dilaton field,  i.e.,  $\Phi'(x)$ or $\Sigma'(x)$.  What the foregoing discussion implies is that we will always find that $\Phi'(x)$ or $\Sigma'(x)$ would diverge for a finite value of $x$.  Clearly,  so would $\Phi$ or $\Sigma$.   We have chosen our parameters in such a way  that it is ensured that effective Newton constant becomes large.  The curvature singularity,  if any,  is present precisely on this surface.

Does the inclusion of matter improve things? In the previous examples \eqref{phidp1} and \eqref{scssm1}, we see that the matter contributes to the dynamical equation with the \emph{same sign} as the extant quadratic term.  
This implies that the finite-time blow-up occurs even more ``quickly'' (i.e.,  at a value $x>0$) in the presence of matter. 

The coupled differential equation for the matter-dilaton system is of course rather hard to solve  in general.   Let us show the reduction of the duration using a simpler system. We add a small perturbation $\Delta^2$ to the equation \eqref{sys1}  (see Figure \ref{fig:dynsysc})
\baa
\dot{X} = X^2 + \Delta^2. \label{sys3}
\ea
Let us solve the equation with the same initial condition as before,  $X(0) = \epsilon$.  Let us assume $\Delta^2 \ll \epsilon^2$ so that the perturbation is really small.  The solution to the equation with this initial condition is,
\baa
X(t) = \Delta  \tan \left(\Delta t+  \tan ^{-1}\left(\frac{\epsilon }{\Delta }\right)\right),  \label{solsys3}
\ea
which blows up at,
\baa
t &= \frac{1}{\Delta} \pqty{ \pm  \frac{\pi}{2} - \tan^{-1}\qty( \frac{\epsilon}{\Delta} )  } \\
&= \frac{1}{\epsilon} \pqty{  1 - \frac{\Delta^2}{3 \epsilon^2} + \mathcal{O} (\Delta^4/\epsilon^4) }, \label{tbluv}
\ea
where the upper (lower) sign is to be chosen for a positive (negative) $\epsilon$.
Therefore,  this system  shows  finite-time blow-up both in the past and the future.  The duration of blow-up is, however, much shorter.

\section{Analytical and Numerical Results}\label{sec-results}

Now that we know that we will encounter singularities generically,  we want to learn  about the gravitational dynamics of the systems at a more detailed level. We shall be using a combination of analytical and numerical techniques.  The goal of this section is to look at certain important qualitative aspects of collapse.  The very first  questions involving gravitational collapse one has in mind concern the nature of the curvature singularity and formation of (apparent) horizons when the singularity is spacelike.   Besides analytical results,  we will present a sampling of numerical results to illustrate the variety of possibilities.   There are several parameters in the problem,  which can give rise to a rather diverse set of results.   One need look no further than Appendix \ref{sec-static} to see an example. 

In discussing the self-similar coordinate system,  we had assumed that $v >0$.  Since we saw that $v=0$ corresponds to a curvature singularity for some geometries,  we will always consider the finite interval $\varepsilon \leq v < \Lambda$,  where $\varepsilon>0$ is a fixed number (recall that we have set the characteristic length scale to unity) and $\Lambda$ is a suitable large number (to avoid any complication from $v\to \infty$ limits).  We will be turning on matter only at $v  = \varepsilon$.  This choice of coordinates describes only a part of the spacetime. We attach the geometry with matter in $v > \varepsilon$ to the vacuum geometry in the region $v < \varepsilon$ and match the two geometries along the null surface $v = \varepsilon$.  Since the scalar field is turned on suddenly at $v=\varepsilon$,  there will be some discontinuity across $v = \varepsilon$,  which can be interpreted as a shock wave.

We would essentially be solving a pair of coupled differential equations (involving the dilaton field and the matter field).   We will need some initial conditions in $x$ to solve the differential equations.  
Physical considerations readily tell us at which value of $x$ we should impose the conditions.  When the asymptotics are Minkowski-like,  the initial conditions should be imposed on the past null infinity $\mathscr{J}^-$,  where the effective gravitational coupling is very weak.   It is easy to see from previous discussions that this happens as $x\to \infty$, for fixed and finite $v$ (this feature is common to all models).   Since we have taken $v \geq \epsilon$,  the $x \to \infty$ limit is not associated with $v \to 0$ and finite $\rho$. For numerical analysis,  it is not possible to impose initial conditions at $x = \infty$.  We will impose them at $x=x_L$, where $x_L$ is some large number and integrate ``back in time''.

Since the boundary of the spacetime is \emph{defined} by the dilaton fields,  we will use the same boundary conditions for these fields and their derivatives as the vacuum solution, namely,
\baa
\Phi(x_L ) &= - \frac{2}{4-\gamma} \log x_L ,  &\quad \Phi(x_L ) &= - \frac{2}{4-\gamma} \frac{1}{x_L}, \\
\Sigma(x_L ) &=  \log x_L ,  &\quad \Sigma'(x_L ) &= \frac{1}{x_L}. \label{gelinc}
\ea
For the stringy model with $\gamma = 4$,  we will do essentially the same thing, with a slight adaptation to suit the numerics.   In this section,  we have set $V_\psi = 0$ throughout,  which corresponds to a free massless scalar in higher dimensions (for the stringy system) or a massless free two-dimensional scalar in two dimensions (for the Liouville system).  For the stringy models, it turns out that even with
\baa
\Psi(x_L) = 0 = \Psi'(x_L), \label{psiinitc}
\ea
non-trivial dynamics is generated due to a non-zero $\lambda$.   Therefore,  we use the condition \eqref{psiinitc} for the stringy systems and vary $\lambda$,  which is a very natural boundary condition on the null infinity.  For the Liouville systems,  we have to have a non-zero value of  $ \Psi'(x_L)$ to start the numerics and we use initial values,
\baa
\Psi'(x_L )  = \epsilon , \quad \Psi (x_L ) =0, \label{psiliointo} 
\ea
to start the numerical integration, where $\epsilon$ is some small non-zero number.

Due to the relative complexity of the $(\rho, v)$ coordinate system,  we will refrain from making comments that depend on this coordinate system and instead make statements that are coordinate-invariant.  
The characterisation of the singularity is not difficult.   If the singular hypersurface is given by $x = x_0$,  then the Ricci scalar generally  blows up in the vicinity of $x=x_0$.  Since we approach this point from the right, we shall find\footnote{Since we have restricted $v \geq \epsilon$,  there is no divergence associated with the $v^2$ in the denominator of the definition of the Ricci scalar \eqref{ricUR}. },
\baa
\qty| F''(x_0^+) | \to \infty. \label{singdef}
\ea

\begin{figure}
\centering
\begin{subfigure}{.33\textwidth}
  \centering
  \includegraphics[height=1.5in]{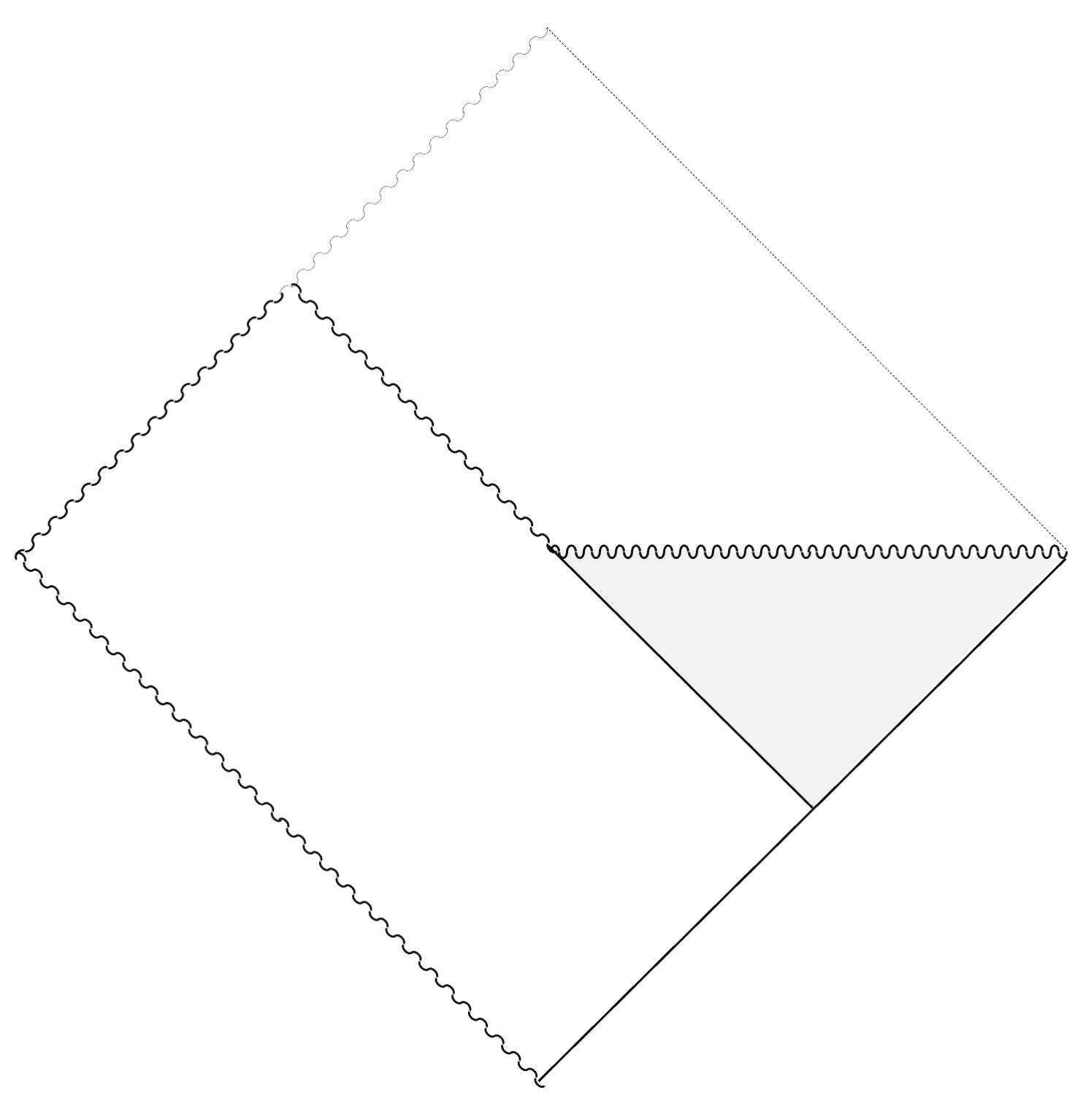}
  \caption{}
  \label{fig:dyna}
\end{subfigure}%
\begin{subfigure}{.33\textwidth}
  \centering
  \includegraphics[height=1.5in]{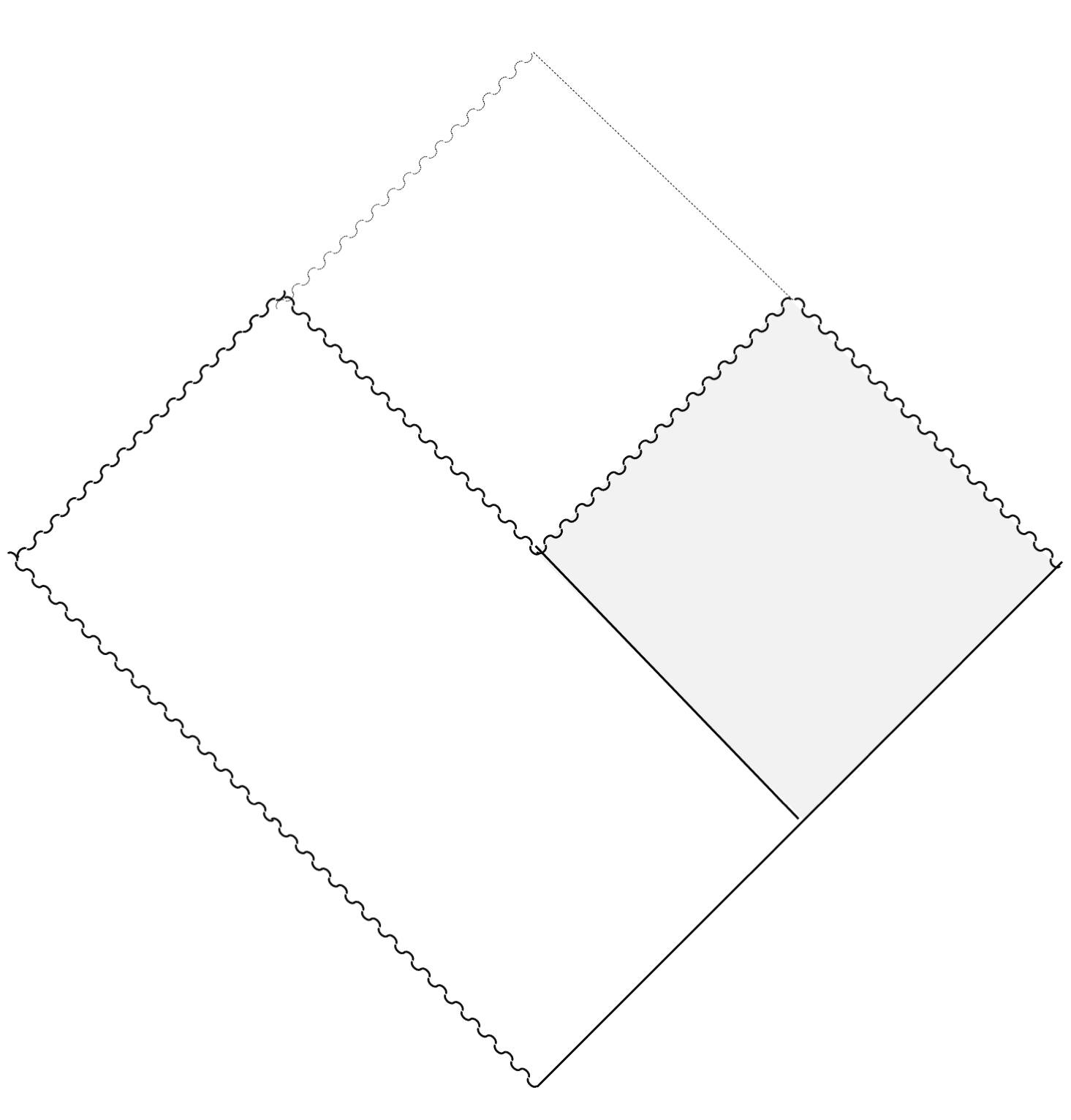}
  \caption{}
  \label{fig:dynb}
\end{subfigure}%
\begin{subfigure}{.165\textwidth}
  \centering
  \includegraphics[height=1.5in]{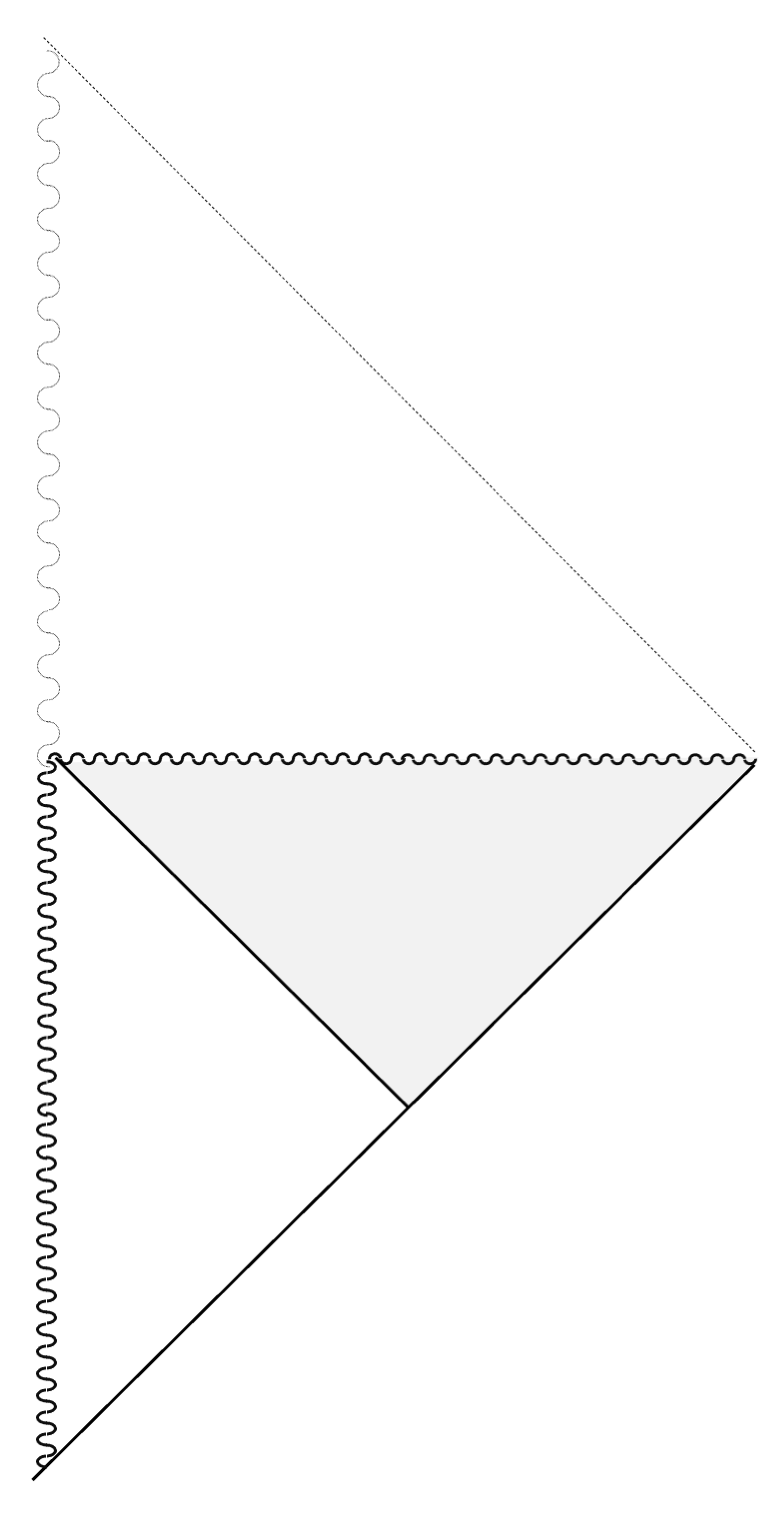}
  \caption{}
  \label{fig:dync}
\end{subfigure}%
\begin{subfigure}{.165\textwidth}
  \centering
  \includegraphics[height=1.5in]{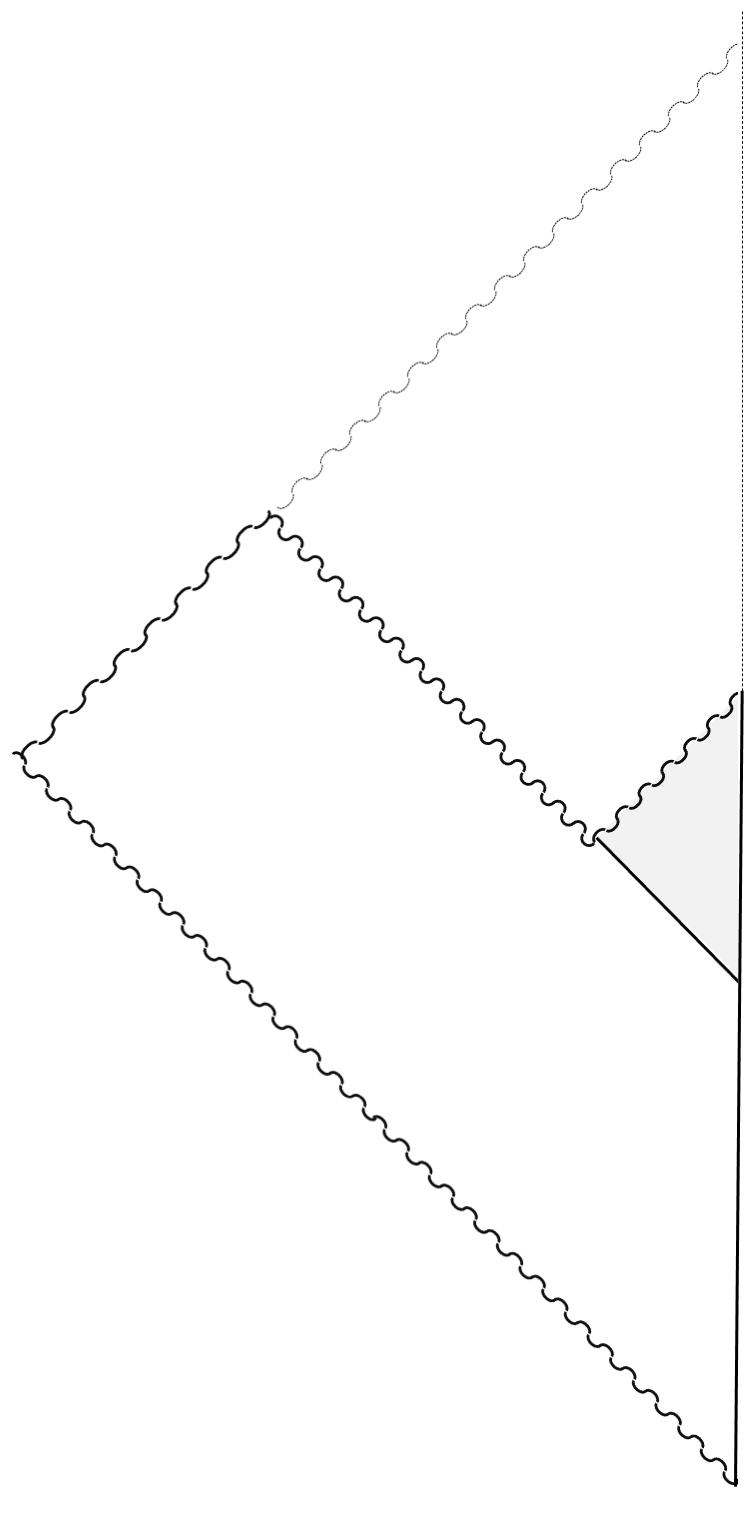}
  \caption{}
  \label{fig:dynd}
\end{subfigure}
\caption{Some qualitatively different possible causal structures under dynamical evolution.  The static spacetime is matched to a dynamical spacetime (grey) along $v = \epsilon$.  The region surrounded by the thin lines does not exist.  {For asymptotically Minkowski-like spacetimes with an initially null singularity,  the matter pulse can result in a spacelike (a) or a shifted null singularity (b).  When the initial singularity (or the strong coupling region) is timelike,  one finds a spacelike singularity (c) under dynamical evolution.  For AdS-like asymptotics with a null singularity,  one finds a shifted null singularity (d). }}
\label{fig:dyn}
\end{figure}

On the other hand the causal nature of the singularity is given simply by the sign of $F(x)$ in the vicinity of $x=x_0$:
\baa
F(x_0^+) \begin{cases}
&>0 \Rightarrow \text{timelike singularity}, \\
&=0 \Rightarrow \text{null singularity}, \\
&<0 \Rightarrow \text{spacelike singularity}. \label{charsin}
\end{cases}
\ea 
The condition above is easily derived from considering the hypersurface where the singularity occurs,
\baa
\frac{\rho}{v} - x_0 =0.  \label{hss}
\ea
The causal nature of this surface is given simply by the sign of the norm of the vector field normal to this surface.  The norm in question is given  by $F(x)/v^2$,  i.e,  in terms of the function that appears in the metric ansatz \eqref{cssan}.  it goes without saying that this result is robust even if the hypersurface were defined by the relation, 
\baa
H(\rho, v) (x - x_0 ) =0,  \label{hhyp}
\ea
where $H(\rho, v)$ is some arbitrary  function (at least $C^1$) of $\rho, v$.
In that case,  there do exist additional contributions to the norm but these vanish as $x \to x_0$.   The norm near the hypersurface \eqref{hhyp} is then given by $F(x_0^+) H(x_0 v, v)^2 / v^2$,  which means that the characterisation \eqref{charsin} holds true no matter what the function $H(\rho, v)$ is.  Based on our results,  we have qualitatively depicted some possible causal structures in Figure \ref{fig:dyn}.
We see that the dynamics of the singularity is sometimes reminiscent of the Dray--'t Hooft geometries \cite{Dray:1984ha}.  {In such geometries,  a high energy shock-wave would be sent towards a black hole and one would then match the geometries across the shock-wave.  The black hole horizon would be shifted relative to its initial position --- we observe a similar phenomenon here vis-\`{a}-vis the singularities.}

We will also attempt to learn whether there exists an apparent horizon in the spacetime,  when there is a spacelike singularity present. In two-dimensional gravity,  in analogy with gravity in higher spacetime dimensions, the locus of the apparent horizon is given by a zero value of the norm of the gradient of the pre-factor of the Ricci scalar in the action (see \cite{Moitra:2019xoj}).  For the stringy system,  for example, the relevant equation is given by
\baa
\qty( \nabla \phi )^2 = \frac{\Phi '(x) \left(F(x) \Phi '(x)-2 \kappa \right)}{v^2} =0 \label{ahdef}.
\ea
The generalisation to the Liouville system is straightforward.  By our boundary conditions,   we have $\Phi'(x \to \infty) \to 0$.  Therefore,  if this condition describes an apparent horizon, it is a trivial one,  at asymptotic infinity.  We call the function,
\baa
G(x) = F(x) \Phi'(x) - 2 \kappa,  \label{ahdef2}
\ea 
a testing function for the non-trivial apparent horizon.

When there is a spacelike singularity and a non-trivial apparent horizon,  an interesting quantity to study would be the quasi-local mass function, which is well-suited for application in dynamical situations such as this.   In previous works such as \cite{Brady:1994xfa,  Aniceto:2019mco},  it was found that the quasi-local mass function evaluated on the apparent horizon exhibits  a Choptuik-like scaling law. We will try to see if such a scaling law exists for the non-trivial apparent horizon.

Before we describe the results, let us make a comment on the numerical methods used.  We argued that the existence of singularity is generally guaranteed  --- this fact also makes the numerics a very difficult exercise.  One has to be extremely precise in the vicinity of the singularity of the solution,   see
\cite{antia2012numerical}. In particular,  one needs to have some form of adaptive step size control to get reliable results around the singularity,  especially when $F(x)$ goes to zero at the singularity.  We have taken this into account and used different methods to compare the results around the singularity to get reliable results.

\subsection{Zero Potential: 2D Phase Space}\label{subsec-zero}

One of the simplest possibilities in the parameter space is when we consider zero values of the potential for both the dilaton and matter fields  $V_\mathrm{eff} = 0= V_\psi $.  Since the variables $\Phi(x)$ (or $\Sigma(x)$) and $\Psi(x)$ (or $X(x)$) do not appear explicitly in the equations of motion,  the phase space is effectively two-dimensional. 

For the Liouville systems with the scalar $\psi$,  the solution can be put in a closed form. To see this,  we write down the ODEs for $\Sigma,  \Psi$. They are given by,
\baa
\Sigma ''(x) &=  -\Sigma '(x)^2-\Psi '(x)^2,  \\
\Psi ''(x) &= -\frac{\Psi '(x) \left(\Sigma '(x)^2+\Psi '(x)^2\right)}{\Sigma '(x)}.  \label{PsiPxxL0}
\ea

Imposing the initial conditions \eqref{gelinc} and \eqref{psiliointo}, we easily find the solution,
\baa
\Sigma'(x) &= \frac{1- x_0/x_L}{x-x_0}, \\
\Psi'(x) &= x_L \epsilon  \frac{1- x_0/x_L}{x-x_0},  \label{PsiPxxL}
\ea
where,
\baa
x_0 = x_L \frac{(x_L \epsilon)^2}{1+(x_L \epsilon)^2}.  \label{x0xLep}
\ea
These derivatives diverge as $x\to x_0$,  the fields $\Sigma,  \Psi$ themselves diverge logarithmically around this point.  Note that if we take  $x_L \to \infty$ and $\epsilon \to 0$ in such a way that $x_L \epsilon^2  =$ finite,  then $x_L- x_0 >0$ and finite.  Therefore,  under this scaling,  the singularity is reached at a shorter distance in $x$.

On account of \eqref{sigric}, we find the Ricci scalar is zero everywhere.  The function $F(x)$ vanishes linearly in $(x-x_0)$. Therefore the hypersurface of strong coupling is null in nature.  Clearly,  there is also no non-trivial apparent horizon in the spacetime.  The analogue of the function $G(x)$ defined previously,  eq.  \eqref{ahdef2},  evaluates to a constant function on the solution.

We get completely analogous results for the semi-classical model,  so we do not repeat the analysis here.  In this case,  the with the boundary condition $X'(x_L) = \epsilon$,  the blow-up occurs at,
\baa
x_0 = \zeta x_L \frac{x_L \epsilon (1+ x_L \epsilon )}{1 + \zeta (x_L \epsilon )^2},  \label{x0zetxl}
\ea
which tells us that the value of $x_0$ becomes independent of $\zeta$ at sufficiently large values of $\zeta$. This is a rather curious result --- adding more effective degrees of freedom does not shift the ``singularity'' around! However,  we have to exercise caution with taking a large $\zeta$ limit,  since taking an arbitrarily large $\zeta$ might lead to inconsistencies \cite{Moitra:2019xoj}.  Interestingly,  we see analytically that for $\epsilon <0$,  the ``singularity'' moves inwards.  For $V_{\mathrm{eff} } = 0$, we get an asymptotically flat spacetime and not the AdS-like asymptotics we discussed for the Liouville system, since that analysis relied on a non-zero value of $V_{\mathrm{eff} }$.

Let us finally discuss the stringy system. In contrast with the Liouville systems discussed above,  there is no closed form solution, but we can make some comments analytically.   Looking at  the expression for $F(x)$ in eq. \eqref{fxsolphi}, one might be tempted to consider the special surface,  
\baa
\kappa  ((8- \gamma) \kappa -4) + \lambda ^2 = 0, \label{falscon}
\ea  
in the parameter space and conclude that $F(x) =  0$ for parameters satisfying this constraint.   This constraint corresponds to a singular behaviour\footnote{Not the sort we are interested in.} in the differential equations.  The RHS of eq.  \eqref{psidp1} blows up when this constraint is met.  Interestingly,  if the inequality \eqref{kapcons} is met, the we can never have \eqref{falscon}.

\begin{figure}
\centering
\includegraphics[scale=0.5]{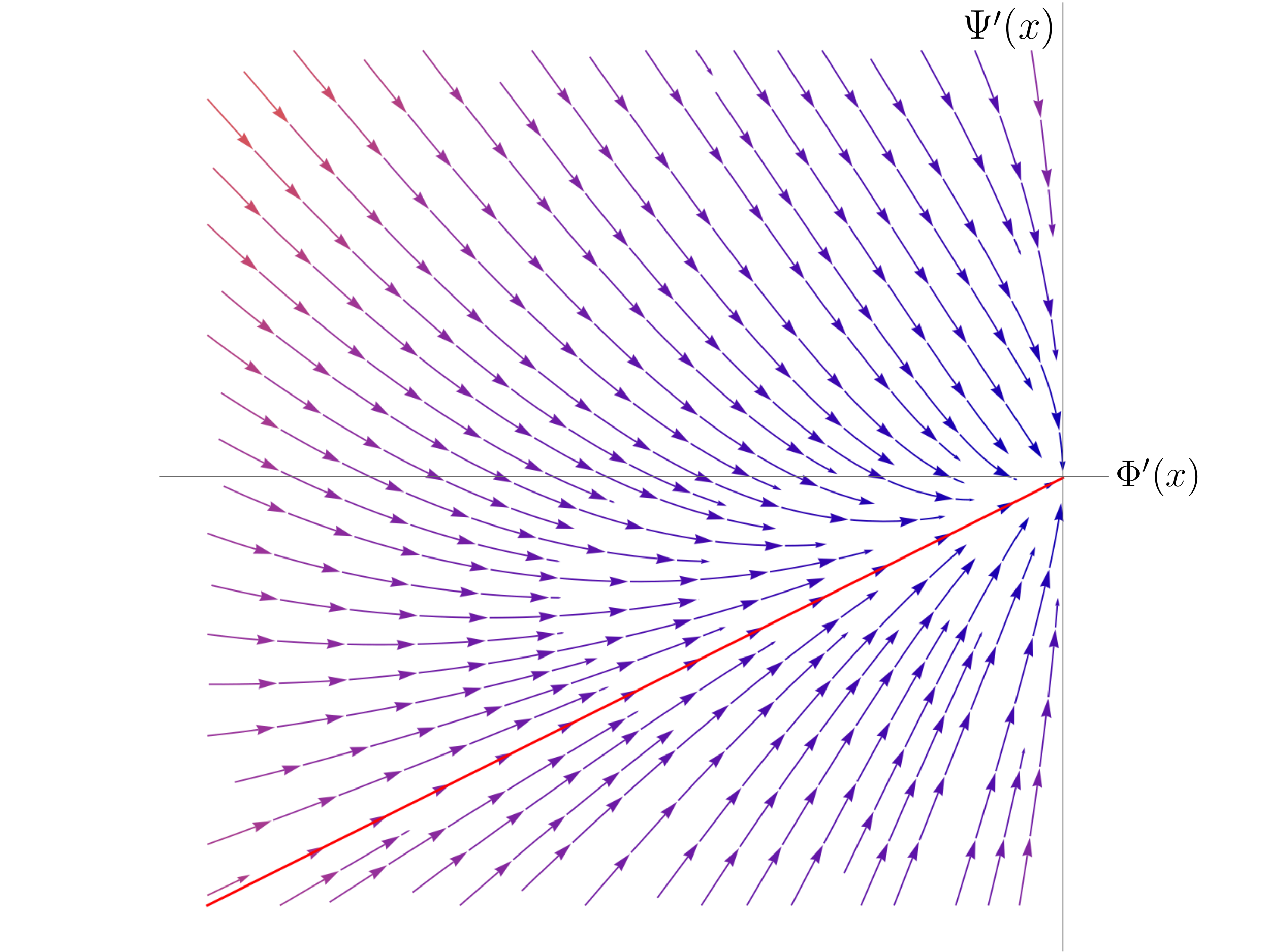}
\caption{The phase plot in the relevant region on the $\Phi'-\Psi'$ plane.  Each curve corresponds to a different solution.  The red curve is the locus of $G(x) = 0$. }
\label{fig:phaseplot}
\end{figure}

We next examine the possibility of formation of a non-trivial apparent horizon, the locus of which is given by,
\baa
-\frac{2 \left(\lambda  \Phi '(x)-\kappa  \Psi '(x)\right)^2}{((\gamma -8) \kappa +4) \Phi '(x)^2+\kappa  \Psi '(x)^2-2 \lambda  \Phi '(x) \Psi '(x)} =0.  \label{nontrivappho}
\ea
This is zero along the surface $\lambda  \Phi '(x)-\kappa  \Psi '(x) =0$. However,  one can show using the differential equations \eqref{phidp1} and \eqref{psidp1} that the relation, $\lambda  \Phi '(x)-\kappa  \Psi '(x) =0$, if true at one value of $x$, will be true for all values of $x$. In other words,  this condition is preserved along the vector field  $(\Phi'', \Psi'')$ in the   $(\Phi', \Psi')$-plane.  A phase portrait is shown in Figure \ref{fig:phaseplot}.  Therefore,  there is no formation of an apparent horizon in this  particular system for any set of values of the parameters.

Let us now describe what information we can glean by a numerical analysis.  Let us note that in the initial geometry,  there is no curvature singularity anywhere and the strong coupling surfaces are null in nature.

For $\gamma = 2$ and $\kappa =1$,  a variation of $\lambda$ gives rise to the following behaviour:  For very small $\lambda$,  say $\lambda =10^{-6}$,  the solution looks very much like the unperturbed solution for a long extent, as one would expect.  It turns out, however, that besides $\Phi(x) \to +\infty$ at a small value $x=x_0$,   there is a spacelike curvature singularity at $x = x_0$.  The curvature singularity was previously absent.  As we dial up $\lambda$,  the singularity becomes much stronger and ``comes outward'', i.e.,  $x_0$ increases. Therefore,  for this choice of parameters,  turning on a profile for the matter field turns the null surface of strong coupling into a genuine spacelike curvature singularity with increasing strength.  Since there is no non-trivial apparent horizon,  we conclude that the singularity stretches to the end of $\mathscr{J}^-$.  See Figure \ref{fig:dyna} for a possible causal structure.

When we try doing the same exercise for larger values of $\kappa$,  i.e.,  $\kappa > 1$ with fixed $\gamma$.  It turns out that from very small values of $\lambda$ up until very large values of $\lambda$,  we get a curvature singularity at $x = x_0$ which is always \emph{null}.  The result indicates that the null singularity after the pulse has moved out.  
The causal structure is given by Figure \ref{fig:dynb}.

One of the questions not addressed by the analysis in \S\ref{sec-interlude} is whether a finite-time blow-up in case of $\gamma = 4$ is guaranteed.  Indeed,  in the absence of matter, one simply has the equation,
\baa
\Phi''(x) = 0,  \label{Phippx0}
\ea
and there is no finite-time blow-up even if we consider the  solution $\Phi'(x) = -a$. ($|\Phi(x)|$  diverges only at $x = \pm \infty$). When one includes matter, the equations are changed and on numerically solving the equations, we find that there is indeed a finite-time blow-up occurring here as well and a spacelike singularity is formed here --- this suggests that the singularity is now a ``true singularity'', reachable in a finite affine time.  (Similar numerical results with $V_{\mathrm{eff}} \neq 0$ are shown in Figure \ref{fig:numc}.)

It is worth pointing out one very interesting feature from  the numerical analysis here.  When $V_{\mathrm{eff} } = 0$,  the strong coupling surface is a null surface in the vacuum for all values of $\kappa$.  When $V_{\mathrm{eff}} \neq 0$,  there is a singularity which is timelike or null depending on the sign of $\kappa - \flatfrac{2}{(4-\gamma)}$.  It is a very interesting that there is a transition in the nature of singularity with  $V_{\mathrm{eff} } = 0$ at precisely this value of $\kappa$! 

Thus,  even in this rather simple setting where we have set all the potentials to zero,  we see rather interesting qualitative behaviour.

\subsection{Numerical Results for Non-Zero Effective Potential}\label{subsec-numst}

We first look at the stringy system.  We now consider the case where $V_{\mathrm{eff}} >0$ and try to answer the same question as before  (however,  we continue to have $V_\psi = 0$). One could observe several important departures from the previous subsection.  For one,  the function $G(x)$ can now smoothly cross zero,  which could signify the existence of an apparent horizon. Turning on $V_{\mathrm{eff}}$ springs some surprises and we get rather interesting results.  

\begin{figure}
\centering
\begin{subfigure}{.50\textwidth}
  \centering
  \includegraphics[width=\linewidth]{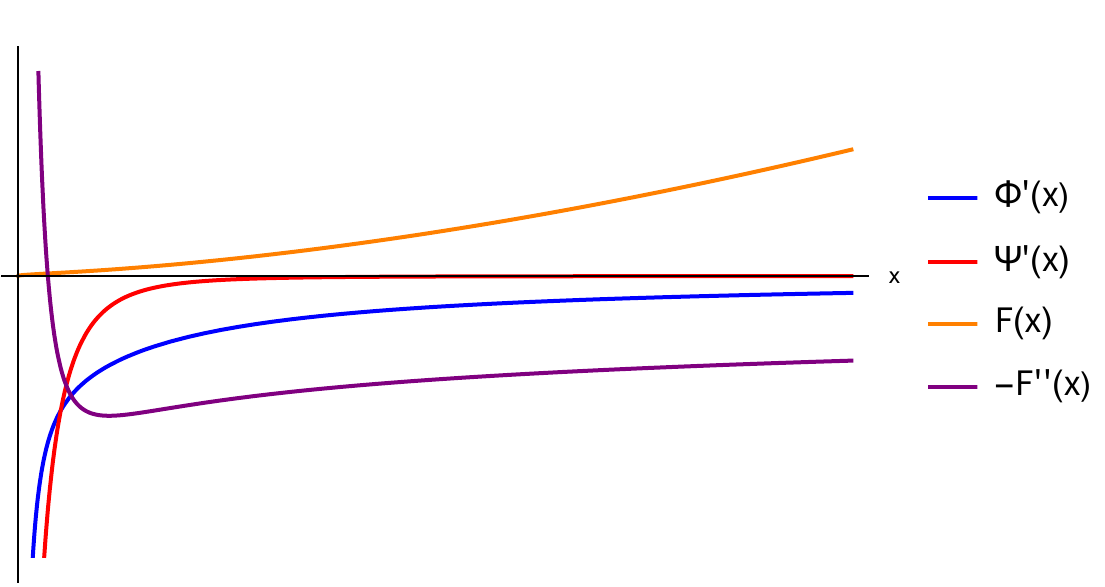}
  \caption{}
  \label{fig:numa}
\end{subfigure}%
\begin{subfigure}{.50\textwidth}
  \centering
  \includegraphics[width=\linewidth]{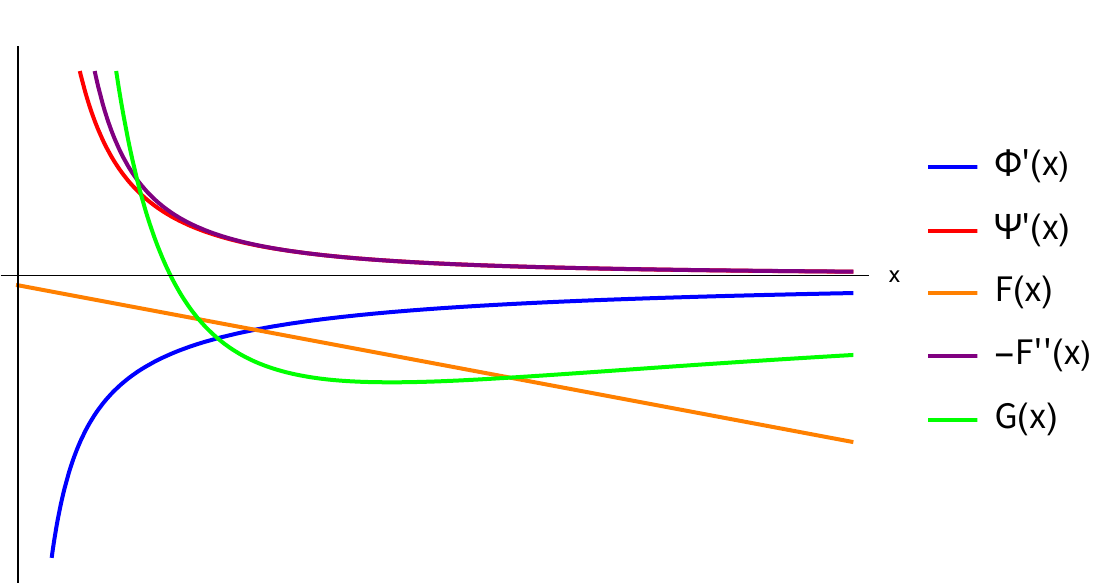}
  \caption{}
  \label{fig:numb}
\end{subfigure}\\
\begin{subfigure}{.50\textwidth}
  \centering
  \includegraphics[width=\linewidth]{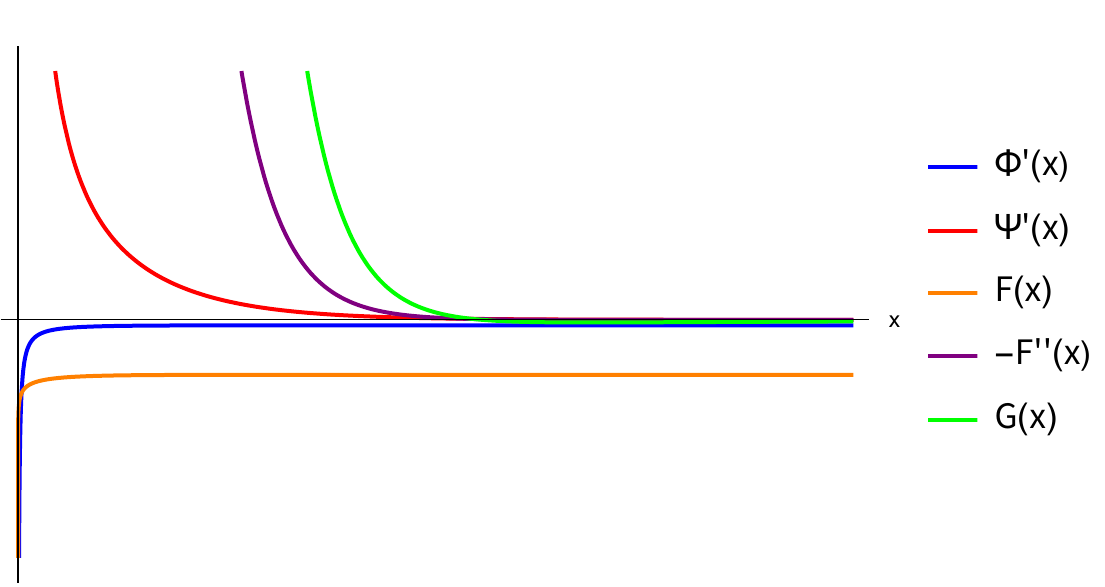}
  \caption{}
  \label{fig:numc}
\end{subfigure}%
\begin{subfigure}{.50\textwidth}
  \centering
  \includegraphics[width=\linewidth]{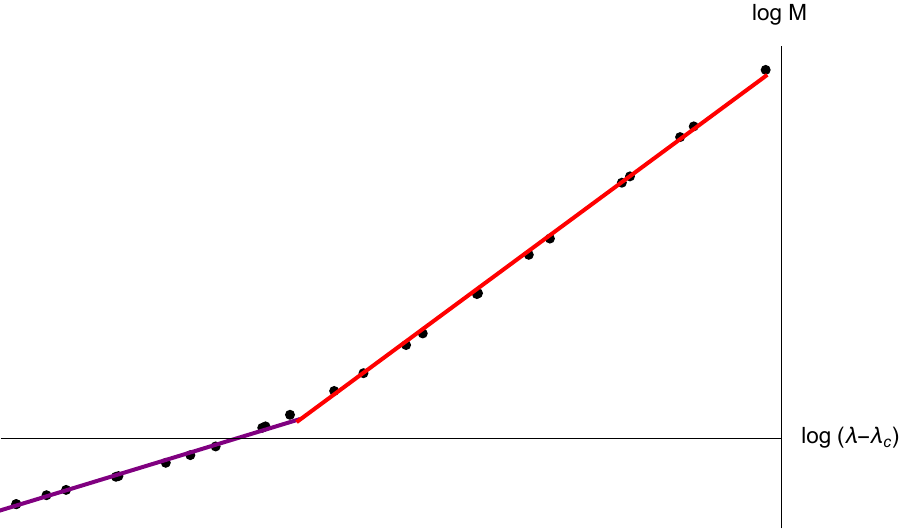}
  \caption{}
  \label{fig:numd}
\end{subfigure}
\caption{A sample of the numerical results.  In the first three plots,  various quantities are plotted against $x$ in the vicinity of the singularity at $x=x_0$ (the origin).  For all the plots $V_{\mathrm{eff}} =1$ and $x_L = 10^6$.  For (a),  $\gamma=2$,  $\kappa= 6$,  $\lambda=10^{-4}$;  for (b), $\gamma=2$,  $\kappa=1$,  $\lambda=10^{-6}$,  and for (c) $\gamma=4$,  $\kappa=10$,  $\lambda = 10^{-10}$, $a=10^{-4}$.   Many quantities diverge near $x=x_0$.   In (d),  a log-log plot of $M(x)$  (at the apparent horizon) vs $(\lambda- \lambda_c)$ is shown,  with two piecewise linear fits. The parameters for (d) are $\gamma=2$ and $\kappa=1$.}
\label{fig:num}
\end{figure}

At first, we take a set of parameters for which the singularity in the vacuum solution is null.
To be concrete,  we study the cases where the asymptotic structure is AdS-like,  i.e.,  $\gamma =2$,  $\kappa>2$  (the numerical plots for the case $\kappa =6$ case is shown in Figure \ref{fig:numa}).  We choose the initial value at $x_L = 10^6$.  We find that analogous to the $V_{\mathrm{eff}} = 0$ situation,  there is always a null singularity for all the values of $\lambda$ that we have used.  This structure persists for reasonably large values of $\kappa$ with a fixed $\gamma$.   The causal structure looks like Fig \ref{fig:dynd}.  For $\gamma =2$ and  the range $1< \kappa \leq 2$,  the asymptotic structure is Minkowski-like.  However,   we see that the singularity continues to be null in character.   The function $F(x)$ for these cases decays to zero near the singularity.

We next turn our discussion to the special values of parameters $\gamma =2$,  $\kappa =1$, the values of parameters related to the Schwarzschild-like black hole \eqref{frgl4}.   The initial geometry has zero curvature everywhere.
For small enough values of $\lambda$,  there appears to be a null singularity.   As we dial up $\lambda$,  above a certain critical value of $\lambda = \lambda_c$, we find that the singularity is spacelike (see Figure \ref{fig:dync}). Furthermore,  we find that the function $G(x)$ does cross zero.  In other words,  there seems to be a non-trivial apparent horizon in the geometry when  we find the spacelike singularity.  The quasi-local mass function $M(x)$ (suitably multiplied with some power of $v$) defined in Appendix \ref{app-qlm} is useful now.  One naturally wonders whether there is some Choptuik-like scaling phenomenon taking place here.  As mentioned before,  a possible form of scaling \cite{Brady:1994xfa,  Aniceto:2019mco} is the scaling of the mass function $M(x)$ at the location of the apparent horizon (i.e.,  the zero of $G(x)$) with $\lambda$.

Plotting $\log M$ against $\log(\lambda - \lambda_c)$, we see a rather interesting behaviour,  as shown in Figure \ref{fig:numd}.   A straight line fits the numerical data beautifully in the vicinity of $\lambda = \lambda_c$.   As one moves further away from  this value of $\lambda$,  there is a new linear slope governing the mass. This suggests that $M(x)$ is governed by different scaling exponents:
\baa
M \sim c_1 (\lambda - \lambda_c)^\alpha + c_2 (\lambda - \lambda_c)^\beta. \label{mscalalphbt}
\ea
Numerically we find $\alpha \approx 0.8$ and $\beta \approx 1.9$.
There is some transition taking place around the point where these two exponents exchange dominance.   One naturally wonders whether the exponents etc.  are dependent on the regulator $x_L$.  We find that although $\lambda_c$ does depend on $x_L$, there does not seem to be a strong dependence of the exponents on $x_L$.  In fact,  $\lambda_c$ seems to scale inversely with $x_L$ --- thus,  when we take $x_L \to \infty$,  $\lambda_c \to 0$. Therefore,  turning on even a small $\lambda$ would lead to a spacelike singularity with a black hole-like geometry.  The results above are strongly reminiscent of Choptuik-like criticality,  with the critical parameter being $\lambda$.

There are more surprises.  The results above suggest that as we increase $\lambda$ further,  $M$ will also increase. However,  there comes a point for sufficiently large $\lambda$ at which the non-trivial apparent horizon is not formed. In this case,  only the trivial apparent horizon at infinity survives.  In fact,  the non-trivial apparent horizon appears to be an artefact of finite $x_L$.  When we increase $x_L$ by some factor $a$,  the location of the  apparent horizon increases by a much bigger factor --- this suggests that this non-trivial apparent horizon might not survive in the $x_L \to \infty$ limit.   However,  even on increasing $x_L$ by four orders of magnitude,  we find roughly the same exponents $\alpha$ and $\beta$,  which indicates these exponents might actually be universal. 

Some of these results also seem to hold true when we vary $\kappa$ slightly so the initial geometry has an timelike singularity --- this timelike singularity turns out to be unstable and collapses to a spacelike singularity,  with the causal structure given by Figure \ref{fig:dync}.  Analogous non-trivial apparent horizons are seen in this case as well.

In all the cases discussed so far,  the initial condition $\Psi'(x) = 0$ is consistent with a positive mass function at $x = x_L$.  In fact,  the mass function behaves as
\baa
M(x_L) \sim x_L^{4/(4-\gamma)} \lambda^2. \label{mscaltwo}
\ea
Therefore,  as we take the limit $x_L \to \infty$, keeping $\lambda$ fixed,  the mass function diverges.  This fact is not too surprising in the context of self-similar collapse.  In situations such as this,  one has to normalise various quantities appropriately to extract physically meaningful quantities.  Put this way,  we can perhaps interpret the above as a Choptuik-like scaling in $\lambda$ with exponent being given by the integer 2! In fact,  $\beta$ is quite close to 2. It would be interesting to examine the scaling behaviour further.

However,  the issue of defining quasi-local mass in general relativity is extremely subtle  --- more so when there is additional matter present. We have used one proposed definition from the literature. It is conceivable that other (inequivalent) proposals of mass function could be of more direct utility for the problem at hand. We leave such a possibility for a future study.

We finally study the case of stringy coupling,  $\gamma =4$.  In this case, for the numerics to be better controlled, we choose the following initial conditions,  $\Phi(x_L) = - a x_L$ and $\Phi'(x_L) = -a$,  where $a$ is a small number.  For the data presented  in Figure \ref{fig:numc},  we have chosen $a = 10^{-4}$,  $\lambda =10^{-10}$ and $\kappa = 10$.  In this case,  we find a strong spacelike singularity accompanied by an apparent horizon.  This is very similar to the $V_{\mathrm{eff} } = 0$ case.
It is worth mentioning that if we take $a$ to be very small keeping all the other parameters fixed,  we find a very strong \emph{timelike} singularity.  If this transition between the nature of the singularities is not some finite cut-off or numerical artefact and has some genuine physical origin,  it would be interesting to examine it.  We hope to return to this question in a future work. It is also worth mentioning that for $V_{\mathrm{eff} } = 0$,  the singularity always seems to be spacelike,   no matter how small $a$ is.

We end this section with remarks on the numerical results for the Liouville system with $V_{\mathrm{eff}} \neq 0 = V_\psi$.
Using allowed values of the parameters $\kappa$,  we found that for a wide sampling of values of $\Psi'(x_L)$ and $\lambda$,  we get a null singularity, which has moved ``outward''.   Therefore,  the causal diagram is that of Figure \ref{fig:dynd}.
What happens when we study the semi-classical $\chi$ system? Again,  the results are not very different from what is observed in the Liouville system with  the minimal scalar.  We always form a null singularity.  

In summary,  we find very interesting behaviour in all corners of the parameter space for the models we considered.

\section{Discussion and Future Directions}\label{sec-concl}

The investigation reported in this article clarified some aspects of self-similar gravitational collapse and nature of singularities in two dimensions.  One can think of several ways to ask physically relevant question in related contexts.

Our starting point was the existence of a homothetic Killing vector.  We found a rather wide class of theories consistent with this symmetry.  It would be interesting to see what happens when one relaxes this assumption --- we might see even richer behaviour than has been found in this article.  {It is worth noting that the original Choptuik scaling result was associated with a discrete self-similarity (DSS).  While the evolution equations were ODEs with the assumption of CSS ansatz, it would be interesting to see whether a solution to the  partial differential equations in these systems shows some behaviour pertaining to  DSS.} It is conceivable that other forms of the dilaton potential might be useful in this regard.    Furthermore,  we imposed some constraints on the parameters of the theory seeking some nice properties of the static spacetimes.  Relaxing these constraints might prove to be useful, especially when one considers microscopic realisations to model self-similar collapse.    It would also be interesting to see if variations of the matter content lead to results of similar nature.

We worked in a particularly simple coordinate system,  in which the ODEs were autonomous --- which made the resulting analysis easy.  It would be interesting to see if these features can be extended to some other coordinate systems which are  better suited to answer some specific questions.
{As mentioned in the introduction,   it would be interesting to use the results in this article to understand similar questions in higher dimensional scenarios. Theories like the Jackiw-Teitelboim model \cite{Teitelboim:1983ux,  Jackiw:1984je} arise naturally in the near-horizon region of near-extremal black holes. In the set-ups of our consideration,  there were singularities instead of black holes. With appropriate ingredients, one could construct analogous singularities in higher dimensional set-ups. It would be interesting to explore whether, in the relevant limit, aspects of the dynamics of such singularities are captured by the set-up explored in this article or deformations thereof.}

Only a selected sample of extensive numerical investigations has been presented in Section \ref{sec-results} of this article.   Many of the results gleaned from the numerical analysis are quite intriguing. For example,  we found that a timelike singularity if present initially,  quickly disappears and becomes a spacelike singularity stretching to null infinity.  This is a very pleasing result from the point of view of the weak cosmic censorship conjecture,  and it would be nice to understand this result from a microscopic perspective.   We got a tantalising hint of Choptuik-like criticality in our numerical results --- it would be interesting to explore this aspect in a wider region of the parameter space.  Another very interesting result was the fact that whenever the asymptotic causal structure resembled AdS asymptotics,  the singularity  in the dynamical situation was null.   

Besides addressing some of the questions raised above,   we also hope to present some more detailed aspects of the numerical analysis in future works.  

\acknowledgments

I thank the ICTP for support. I also thank the following organisations for their hospitality during the progress of this work: the Max Planck Institute for the Physics of Complex Systems,  the Simons Center for Geometry and Physics,  Stony Brook University, (during the 2022 Simons Summer Workshop),  the University of Chicago, the University of Wisconsin--Madison and the University of Michigan, Ann Arbor. 

\appendix

\section{Static Black Hole Solutions}\label{sec-static}

In Section \ref{sec-selsi}, we set-up the stage for discussing self-similar solutions in a detailed manner.  In this Appendix,  we explore something simpler which arises from the actions derived in Section \ref{sec-self} -- static solutions without additional matter.  We will construct black hole solutions in these models with a curvature singularity behind the event horizon.  These static solutions are not self-similar in general.
We explore these solutions to build more intuition about the spacetimes under consideration.  We will examine the nature of the asymptotic structure of these spacetimes and also constrain the parameters that appear in the Lagrangian.  Some related solutions are well studied in the literature --- see the early works \cite{Lechtenfeld:1992rt, Mann:1993rf} where the focus was on finding asymptotically flat black holes.

To find out the static solutions,  we will consider a coordinate system in which the metric is put in the Schwarzschild gauge,
\baa
\dd s^2 = - f(r ) \, \dd t^2 +  \frac{\dd r^2}{f(r)}, \label{metrican}
\ea 
and the dilaton too depends only on the spatial coordinate $r$,
\baa
\phi = \phi (r). \label{dilphir}
\ea
(When we consider the Liouville model,  we will similarly assume that $\sigma = \sigma (r)$).  The vector field $\partial/\partial t$ is a Killing vector of the gravity-dilaton system.  In this coordinate system the expression for the Ricci scalar is particularly simple,
\baa
R  = - f''(r). \label{riccifppr}
\ea
We therefore carry out the usual steps of inserting these ans\"{a}tze in the equations of motion and solving for the functions $f(r)$ and $\phi(r)$ (or $\sigma(r)$).  
\subsection{Stringy System}\label{subsec-ststr}

We find that the off-diagonal $(tr)$ component of the metric equations of motion \eqref{meom2} is identically zero.  A certain combination of the $rr$ and $tt$ components yields the simple differential equation for the dilaton,
\baa
2 \phi ''(r) + (\gamma -4) \phi '(r)^2 = 0, \label{dilatsim1}
\ea
which immediately yields the solution (for $\gamma \neq 4$),
\baa
\phi (r)  =  \phi_0 - \frac{2}{4-\gamma} \log (r-r_0). \label{dilatsolr1}
\ea
We fix the two constants of integration $\phi_0$ and $r_0$ as follows: $\phi_0$ can be set to zero by a redefinition of the 
 effective potential strength $V_{\mathrm{eff}}$ (at the level of the equations of motion), 
 while on the other hand,  $r_0$ simply corresponds to the ``origin'' of the $r$ coordinate and can thus be set to zero.  In other words,  we take the $r$ coordinate to lie in the range $0 < r <\infty$.  A characteristic length scale appears within the logarithm to make its argument dimensionless -- we have conveniently set this length scale to unity.   Finally,  the dilaton takes the form,
\baa
\phi (r)  = - \frac{2}{4-\gamma} \log r.  \label{phisolog}
\ea
The dilaton can thus assume any real value. When $\gamma < 4$,  the dilaton $\phi$ monotonically decreases from $+\infty$ to $-\infty$  as a function of $r$.   Note that the Ricci scalar couples to,
\baa
\exp(-2\phi)  = r^{ \frac{4}{4-\gamma}  }. \label{ricpref}
\ea
Therefore,  for identification of $r$ with the conventional radial coordinate (where the asymptotic region is located for large values of $r$),   we shall take, 
\baa
\gamma < 4, \label{gamrange}
\ea 
(we will treat the ``stringy coupling'' $\gamma =4$  separately).   We will also impose the constraint $\gamma  \geq 0$, which is relevant for generic dimensional reductions.

Let us now solve for $f$. We obtain from the $rr$ component of \eqref{meom2} the following first order equation in $f$,
\baa
2 f'(r) \phi '(r)-\gamma  f(r) \phi '(r)^2+V_{\mathrm{eff}} e^{\frac{2 \phi (r)}{\kappa }} = 0. \label{freq1}
\ea
The solution is given by,
\begin{equation}
f(r) =  \frac{(4-\gamma)^2 \kappa  V_{\mathrm{eff}}}{4 ((8 - \gamma) \kappa -4)} r^{2 - \frac{4}{(4-\gamma ) \kappa } } \bqty{1 - \pqty{\frac{r}{r_h} }^{\frac{\gamma  \kappa -8 \kappa +4}{(4-\gamma ) \kappa }}  }, \label{bfphi1}
\end{equation}
where $r_h$ is a constant of integration, carefully chosen so that the event horizon of the black hole lies at $r=r_h >0$.

We must have $f(r) > 0$ outside the black hole,  i.e., when $r>r_h$.  This requirement implies, in  conjunction with the condition \eqref{gamrange} that,
\baa
V_{\mathrm{eff}} > 0. \label{veffgr}
\ea
We can find the form of Ricci scalar from eq.  \eqref{riccifppr},
\baa
R = ( \cdots ) r^{-\frac{4}{(4-\gamma) \kappa }} + (\cdots) r^{ - \frac{8-\gamma }{4-\gamma } }. \label{starexp}
\ea
Since we want our asymptotic region to lie near $r \to \infty$ and the singularity at $r=0$, we require $\kappa >0$.   Note that with this constraint, $R\to 0$ as $r\to \infty$ and $|R| \to \infty$ as $r\to 0$.  Therefore,  the asymptotic region $r\to \infty$ has zero curvature and in the region near $r=0$, the curvature becomes arbitrarily large. Whenever $r_h > 0$,  this (spacelike) singularity is shielded by the event horizon -- there is a null or  timelike singularity otherwise.  The parameter $r_h$ can be related to the mass carried by the black hole (see Appendix \ref{app-qlm}).   It so happens that  we actually have a stronger constraint on $\kappa$,
\begin{equation}
\kappa > \frac{4}{8-\gamma} .  \label{kapcons}
\end{equation}
We can derive this constraint as follows. We want the  function $f(r)$,  eq.  \eqref{bfphi1},  to be positive everywhere outside of the black hole $r>r_h$.   For all positive values of $\kappa$ (except for the singular point $\kappa = 4/(8-\gamma)$),  the function $f(r)$ is positive.   However,  since we shall identify some positive power of $r_h$ with the (quasi-local) mass of the black hole,  if we want the $r_h \to 0$ limit to be smooth,  we easily see from \eqref{bfphi1} that we need the constraint \eqref{kapcons}. (See Appendix \ref{app-qlm})

At large values of $r$,  the function $f(r)$ may either decay to zero,  or grow unboundedly or assume a constant value depending on the leading exponent of $r$ in $f(r)$, i.e.,  $2 -  4/(4-\gamma)\kappa$.  Although the curvature is vanishingly small in the asymptotic region,  the asymptotic causal structure depends on the value of $\kappa$.

When $\kappa < \flatfrac{2}{(4-\gamma)}$,  the leading power of $r$ is negative and the function $f(r)$ decays to zero at large $r$.   
The asymptotic causal structure for this range of $\kappa$ is similar to that of Minkowski spacetime.  A more careful analysis reveals that this causal structure persists till the value $\kappa \leq  \flatfrac{4}{(4-\gamma)}$.  For the regime $\kappa  >  \flatfrac{4}{(4-\gamma)}$, the asymptotic causal structure is similar to that of the AdS spacetime,  even though the curvature vanishes in this region. As mentioned in the Introduction,  spacetimes with such exotic asymptotics have been noted previously in case of higher dimensional dilatonic black holes \cite{Chan:1995fr}.  In two dimensions,  the dilaton and ``gravity'' are inseparable entities --- it is somewhat of a surprise that such 2D spacetimes did not find much  (if any) emphasis in the literature.

Let us clarify the aforementioned discussion with some familiar examples for some special values of $\kappa$.   For $\kappa =  \flatfrac{2}{(4-\gamma)}$,  for instance,  the function $f(r)$ is given by,
\baa
f(r) = \frac{(4-\gamma)^2}{4\gamma} V_{\mathrm{eff}} \bqty{ 1 - \pqty{ \frac{r_h}{r} }^{\frac{\gamma}{4-\gamma}}  },  \label{frgl4}
\ea
which shows that for this value of $\kappa$,  the metric in this coordinate system assumes the form similar to the Schwarzschild black hole. For example,  for a $d$-dimensional Schwarzschild black hole,  $\gamma$ would correspond to $\gamma = \flatfrac{4 (d-3)}{(d-2)}$.  To see how this value of $\gamma$ comes about,  one can consider a general dimensional reduction for a spherically symmetric black hole.  See, for example,  \cite{Moitra:2021eom}, eq.  (2.255), which yields this value of $\gamma$ after appropriate field redefinitions.  Note that this value of $\gamma$ is always less than 4 (the string theory coupling) and reaches this value only as $d \to \infty$.

Another well-known example is the borderline case $\kappa =  \flatfrac{4}{(4-\gamma)}$,
\baa
f(r) = \frac{(4-\gamma)^2}{16} V_{\mathrm{eff}}  \, r \, \bqty{ 1 - \pqty{ \frac{r_h}{r} }^{\frac{4}{4-\gamma}}  }.
\ea
For large $r$, $f(r)$ grows linearly in $r$,  which is just the Rindler metric in disguise.  Therefore, the asymptotic causal structure is that of Minkowski spacetime.

On the other hand, the limit $\kappa \to \infty$ illustrates the rather interesting nature  of the asymptotic behaviour for sufficiently large values of $\kappa$.   In this case,   the function $f(r)$ grows quadratically in $r$ for large $r$,  as in asymptotically AdS spacetimes,
\baa
f(r) \approx \frac{(4-\gamma)^2 V_{\mathrm{eff}} }{4(8-\gamma) } \, r^2 \bqty{  1 - \pqty{\frac{r_h}{r} }^{ \frac{8-\gamma}{4-\gamma} }   }. \label{frrind}
\ea
For a general $\gamma$,  the spacetime described by this function still has a curvature singularity at $r= 0$.  However, at the value $\gamma = 0$,  the spacetime becomes exactly $\mathrm{AdS}_2$,  as it must, because the modified action with $\gamma =0$ and $\kappa = \infty$ is nothing but the Jackiw-Teitelboim model \cite{Teitelboim:1983ux, Jackiw:1984je}.

As mentioned previously,  a non-zero constant value of the scalar curvature,  $R = \Lambda$ is inconsistent with the homothety transformation property of the Ricci scalar,  \eqref{rictrans}.  Therefore,  for studying self-similar collapse in asymptotically AdS-like spacetimes with an HKV,  our models (also see the following Liouville models) seem to be very convenient.  {Such spacetimes are interesting because they are often helpful in realising the holographic correspondence ---  a one-dimensional quantum mechanical theory lives on the time-like conformal boundary of the two-dimensional spacetime.  It would thus be interesting to apply holographic ideas in this scenario,  to understand the dynamics from a quantum-mechanical point of view.}

Before we end this subsection, there is an obvious yet important comment to make.   It is immediately obvious that a constant dilaton $\phi (r) = \phi_0$ is a solution of the equation \eqref{dilatsim1}.  This fact is, however, obscured in the solution \eqref{dilatsolr1} --- we cannot recover the constant solution by taking any smooth limit of the integration constants in \eqref{dilatsolr1}.  This constant solution when inserted in \eqref{freq1} immediately tells us that such a solution is possible only if the potential vanishes: $V_{ \mathrm{eff} } = 0$. Under this condition,  we immediately obtain $R = 0$ everywhere and therefore the solution describes simply flat spacetime.

\subsection{Stringy Coupling $\gamma =4$}
\label{subsec-ststr4}

We solve the system of equations as before,  and we obtain, after fixing an additive constant to zero and setting the constant associated with the characteristic length scale to unity,
\baa
\phi (r) = -  r. \label{gfphi}
\ea
In this case,  however,  in contrast to the case for $\gamma <4$, the range of the coordinate $r$ is given by,
\baa
-\infty < r <\infty.  \label{gfphra}
\ea

The solution is given by,
\baa
f(r) = \frac{\kappa  V_{\mathrm{eff}} }{4 (\kappa -1)} \exp(-\frac{2  r}{\kappa }  ) \bqty{ 1 - \exp(-\frac{2  (\kappa -1) (r-r_h) }{\kappa }  )}, \label{frg4}
\ea
where $r_h$ as before denotes the location of the event horizon. 

The positivity of $f(r)$ outside the event horizon leads to the constraints,
\baa
  V_{\mathrm{eff}} >0,  \quad \kappa >1. \label{gffcon}
\ea
The curvature grows unbounded as one takes $r \to -\infty$.  However, in contrast with the usual singularities in general relativity,  it takes an infinite affine time to reach the ``singularity'' at $r=-\infty$,  so this is not really a singularity \cite{Wald:1984rg}.

In this case,  in contrast with the scenario for $\gamma <4$,   the asymptotic causal structure is the same as that of Minkowski spacetime for all values of $\kappa$.  In particular,  the $\kappa \to \infty$ limit is the black hole discussed in \cite{Mandal:1991tz,  Elitzur:1990ubs, Witten:1991yr} {(which is also the ``pure gravity sector'' of the CGHS model \cite{Callan:1992rs})}.
\subsection{Liouville System}
\label{subsec-stliou}
In this case,  the equations of motion give us, 
\baa
\sigma (r) = \log r. \label{solsigr}
\ea
Since we want the coupling to the Ricci scalar to be non-negative, we restrict $r$ to lie in the range $[e^{-\sigma_0}, \infty)$.   We take $\sigma_0$ (the coefficient of the topological term) to be so large that for all practical purposes (especially while performing numerical analysis),  we take the lower limit of the $r$ coordinate to be zero.  Here,  $r=0$ corresponds to the curvature singularity (the Ricci scalar is exponentially large) and $r=\infty$ is the asymptotic region.  The function $f(r)$ is given by,
\baa
f(r) = \frac{V_{\mathrm{eff}} \kappa }{2+ \kappa } r^{ 2 + \frac{2}{\kappa } } \bqty{   1 - \pqty{\frac{r_h}{r}}^{1 + \frac{2}{\kappa} } }. \label{liofr}
\ea
In this case,  the previously mentioned physical constraints necessitate that,
\baa
V_{\mathrm{eff}} >0, \label{licon1}
\ea
and
\baa
\kappa < -2. \label{licon2}
\ea
The Ricci scalar for this geometry is given by,
\baa
R = - 2 \qty(1 + \frac{1}{\kappa}  )  V_{\mathrm{eff}} r^{2/\kappa }. \label{liric}
\ea 
Clearly,  for finite $\kappa <-2$,  the curvature is exponentially large near $r = 0$ and vanishes at $r=\infty$. 

The asymptotic causal structure is similar to that of AdS spacetime.  Indeed in the limit $\kappa \to -\infty$,  the metric is that of $\mathrm{AdS}_2$ (which of course has no curvature singularity) with the  $\mathrm{AdS}_2$ length $V_{\mathrm{eff}}^{-1/2}$.

\section{Quasi-Local Mass}\label{app-qlm}

We consider the question of quasi-local mass in the two-dimensional theories of gravity discussed in this article.  The authors of \cite{Cai:2016rcv} proposed a definition of quasi-local mass in two-dimensional dilaton gravity in analogy with the Misner-Sharp mass in  higher dimensional theories.  The proposal of \cite{Cai:2016rcv} can be summarised as follows.

If the action describing the gravity dilaton system with matter takes the form,  (setting $16\pi G_2 =1$)
\baa
S= \int \dd[2]{x} \sqrt{-g} \pqty{ \Pi R +  U(\Pi) (\nabla \Pi)^2 + V(\Pi)  } + S_{\mathrm{matter}}, \label{finaca1}
\ea
the proposed definition of quasi-local mass is given by,
\baa
E = \frac12 \pqty{ w - e^Q(\nabla \Pi)^2  },  \label{defqlm} 
\ea
where the functions $Q, w$ are defined by the relations, $U (\Pi) = - Q'(\Pi)$ and  $V(\Pi) = w'(\Pi) \exp[ - Q(\Pi)  ]$.
In certain cases,  we have to modify \eqref{defqlm} by an overall constant of proportionality,  but for the problem we are interested in, this is not an issue.

Comparing with the stringy system,  we find,
\baa
\Pi = e^{-2\phi}, \quad  U(\Pi) = \frac{\gamma}{4\Pi},  \quad Q  = - \frac{\gamma}{4} \log \Pi, \quad W(\Pi) = V_{\mathrm{eff}} \frac{\Pi^{ 2- \frac{1}{\kappa} - \frac{\gamma}{4}}}{2- \frac{1}{\kappa} - \frac{\gamma}{4}} \label{tla1}
\ea
which gives us the form of the mass function in terms of $\phi$
\baa
E = \exp( \frac{\gamma-8}{2}   \phi ) \pqty{ V_{\mathrm{eff}} \frac{e^{2 \phi /\kappa}}{2- \frac{1}{\kappa} - \frac{\gamma}{4}} -  4 (\nabla \phi)^2  }.  \label{tla2}
\ea

For the Liouville system,  we would have,
\baa
\Pi = \sigma, \quad U(\Pi) = -1,  \quad Q (\Pi) = \Pi, \quad w(\Pi ) =  \kappa V_{\mathrm{eff}} \frac{\exp(\frac{2 + \kappa}{\kappa}  \Pi )}{2+ \kappa}, \label{tla3}
\ea
which yields,  
\baa
E =  \frac{\kappa V_{\mathrm{eff}}}{2+ \kappa} \exp(\frac{2 + \kappa}{\kappa}  \sigma ) - \exp(\sigma) (\nabla \sigma)^2. \label{tla4}
\ea
This can be easily generalised to the semi-classical model as well. It can be seen that for the static solutions in Appendix \ref{sec-static},  these are constant and give the correct dependence on $r_h$ (which can be found out from the first law of black hole mechanics) namely,
\baa
E \propto \begin{cases}
r_h^{\frac{(8-\gamma ) \kappa -4}{(4-\gamma) \kappa }} ,  \quad & \text{(stringy model with $\gamma < 4$)}, \\
e^{\frac{2 (\kappa -1) r_h}{\kappa }}\quad & \text{(stringy model with $\gamma = 4$)},  \\
r_h^{1 + 2/\kappa},  \quad & \text{(Liouville model)}.
\end{cases} \label{masscomp}
\ea

In the main text,  for the self-similar system,  we have studied the mass function $M(x)$ defined by taking out an overall power of $v$ in $E$,
\baa
E = v^{-\frac{1}{2} (\gamma -8) \kappa -2}  M(x),  \label{mdef1}
\ea
where,
\baa
M(x) = -\frac{4 e^{\frac{1}{2} (\gamma -8) \Phi (x)} \left(\kappa  V_{\text{eff}} e^{\frac{2 \Phi (x)}{\kappa }}-((\gamma -8) \kappa +4) \Phi '(x) \left(2 \kappa -F(x) \Phi '(x)\right)\right)}{(\gamma -8) \kappa +4}. \label{mdef}
\ea
Analogous relations can be derived for the other models as well.

 \bibliographystyle{JHEP}
 
 \bibliography{refs}

\end{document}